\documentclass[aps,prd,preprintnumbers,superscriptaddress,nofootinbib,twocolumn,10pt]{revtex4-2}
\usepackage{amsmath}
\usepackage{amsfonts}
\usepackage{amssymb}
\usepackage{slashed}
\usepackage[colorlinks, citecolor=blue, linkcolor=blue, urlcolor=blue]{hyperref}
\usepackage{float}
\usepackage[caption=false]{subfig} % Modern alternative to subfigure
\usepackage{tabularx}
\usepackage{pbox}
\usepackage[normalem]{ulem}
\usepackage{orcidlink}
\usepackage{comment}
\usepackage[svgnames]{xcolor}

\def\bal#1\eal{\begin{align}#1\end{align}}
\def\beq#1\eeq{\begin{equation}#1\end{equation}}

\def\GeV{{\rm GeV}}

\def\gagg{g_{a \gamma \gamma}}

\begin{document}
\title{Data-taking strategy for dark-photon dark matter searches with FLASH and BabyIAXO}
%===============================================

\author{Victor Fonoll\,\orcidlink{0009-0003-1740-4474}}
\email{vfonoll@unizar.es}
\affiliation{Centro de Astropartículas y Física de Altas Energías (CAPA), Universidad de Zaragoza, 50009 Zaragoza, Spain}

\author{Maurizio  Giannotti\,\orcidlink{0000-0001-9823-6262}}
\email{mgiannotti@unizar.es}
\affiliation{CAPA \& Departamento de F\'isica Te\'orica, Universidad de Zaragoza, C. Pedro Cerbuna 12, 50009 Zaragoza, Spain}

\author{Giacomo Landini\,\orcidlink{0000-0002-1572-7523}}
\email{giacomo.landini@lnf.infn.it}
\affiliation{Istituto Nazionale di Fisica Nucleare, Laboratori Nazionali di Frascati, \\ C.P.~13, 00044 Frascati, Italy}

\author{Giuseppe Lucente\,\orcidlink{0000-0002-3266-3154}}
\email{lucenteg@slac.stanford.edu}
\affiliation{SLAC National Accelerator Laboratory, Stanford University, Menlo Park, CA 94025}

\author{Federico Mescia\,\orcidlink{0000-0003-3582-2162}}
\email{federico.mescia@lnf.infn.it}
\affiliation{Istituto Nazionale di Fisica Nucleare, Laboratori Nazionali di Frascati, \\ C.P.~13, 00044 Frascati, Italy}

\author{Cristian Cogollos\,\orcidlink{0000-0003-4336-7622}}
\email{cogollos@mpp.mpg.de}
\affiliation{Max-Planck-Institut für Physik (Werner-Heisenberg-Institut), Boltzmannstraße 8, 85748 Garching, Germany}

\author{Igor G. Irastorza\,\orcidlink{0000-0002-4438-4988}}
\email{Igor.Irastorza@cern.ch}
\affiliation{CAPA \& Departamento de F\'isica Te\'orica, Universidad de Zaragoza, C. Pedro Cerbuna 12, 50009 Zaragoza, Spain}

\author{Alessio Rettaroli\,\orcidlink{0000-0001-6080-8843}}
\email{Alessio.Rettaroli@lnf.infn.it}
\affiliation{Istituto Nazionale di Fisica Nucleare, Laboratori Nazionali di Frascati, \\ C.P.~13, 00044 Frascati, Italy}

\author{Simone Tocci\,\orcidlink{0000-0002-5800-5408}}
\email{Simone.Tocci@lnf.infn.it}
\affiliation{Istituto Nazionale di Fisica Nucleare, Laboratori Nazionali di Frascati, \\ C.P.~13, 00044 Frascati, Italy}

%==============================================
\begin{abstract}
The two next-generation axion experiments \href{https://coldlab.lnf.infn.it/experiments/flash/}{FLASH} and  \href{https://iaxo.desy.de/iaxo/}{BabyIAXO} will soon start searching for axion dark matter around $\mu\text{eV}$, at the INFN Laboratori Nazionali di Frascati (LNF) and Deutsches Elektronen-Synchrotron (DESY), respectively.
Both are planning a search for dark-photon dark matter.
If the dark-photon polarization is fixed, its projection on the cavity mode, and hence the signal, changes with the rotation of the Earth.
The published dark-photon projections of both experiments do not exploit this modulation and, for FLASH, treat the unknown polarization in an overly conservative way.
We define the data-taking and analysis strategy for these searches and correct the projections.
Measuring each frequency at three or more sidereal phases, equally spaced over a sidereal day, and orienting the cavity mode, where possible, at $54.7^\circ$ from the Earth's rotation axis recovers the sensitivity of a randomly polarized dark photon, without hardware changes and without affecting the axion search.
The expected sensitivity to the kinetic mixing improves by about \emph{one order of magnitude} for FLASH and by \emph{a factor of four} for BabyIAXO.
\end{abstract}
\preprint{xx}
\maketitle

\section{Introduction and Theoretical Framework}

Many extensions of the Standard Model (SM) predict new light particles with very weak couplings to ordinary matter, collectively known as weakly interacting slim particles (WISPs)~\cite{Jaeckel:2010ni,Ringwald:2012hr,Agrawal:2021dbo,Antel:2023hkf,Arza:2026rsl}.
The best-known example is the QCD axion~\cite{Peccei:1977hh,Weinberg:1977ma,Wilczek:1977pj}, but axion-like particles and hidden $U(1)$ gauge bosons have been 
{the subject of intense experimental and theoretical work in recent years}. {WISPs appear in many extensions of the SM and arise generically} in string compactifications~\cite{Abel:2008ai,Goodsell:2009xc,Ringwald:2012hr,Coudarchet:2025dfd}.
If produced non-thermally in the early Universe, they can account for the dark matter (DM), in the form of a cold classical field with a large occupation number~\cite{Arias:2012az}.
This possibility motivates the broad experimental program, reviewed in Refs.~\cite{Essig:2013lka,Agrawal:2021dbo,Antel:2023hkf,Antypas:2022asj,Adams:2022pbo}.

\begin{table*}[t]
\begin{ruledtabular}
\begin{tabular}{lcc}
    & \textbf{FLASH}
    & \textbf{BabyIAXO} \\
\textbf{Data-taking strategy}    & \textbf{(axis at $57^\circ$)}
    & \textbf{(RADES Mode~1)} \\
\noalign{\smallskip}
\colrule
\noalign{\smallskip}
 & \multicolumn{2}{c}{\textbf{Gain in reach}
   $\mathbf{\chi_{\rm lim}^{\rm published}/\chi_{\rm lim}}$} \\
   \noalign{\smallskip}
\colrule
\noalign{\smallskip}
Published projection
    & $1$   & $1$    \\[3pt]
\quad $+$ Joint statistics (same scan)
    & $2.8$ & $1.13$ \\[3pt]
\quad $+$ {$N\geq$}3 \ sidereal phases per frequency
    & $9.8$ & $3.7$  \\[3pt]
\quad $+$ Best pointing (BabyIAXO only)
    & ---   & $4.2$  \\
\noalign{\smallskip}
\end{tabular}
\end{ruledtabular}
\caption{Gain in the expected $95\%$ C.L.\ exclusion limit on the kinetic
mixing $\chi$ with respect to the published DP projections of
FLASH~\cite{Alesini:2023qed} and BabyIAXO~\cite{Ahyoune:2023gfw}, for DP
dark matter with a fixed, unknown polarization.
A gain of $10$ means a limit on $\chi$ ten times smaller.
Each row adds one step to the previous one, with the same total measurement
time per frequency and no change in the hardware.
\textit{Published projection:} the published projections for a fixed
polarization, which recast the axion projections with
$\langle\cos^2\theta\rangle=0.0025$ (FLASH) and $0.019$ (BabyIAXO).
\textit{Joint statistics:} the same scans, with the noise and the unknown
polarization combined in a single $95\%$ C.L.\ requirement
(Sec.~\ref{sec:signal}); this corrects the published analyses and requires
no change in the data taking.
{\textit{$N\geq3$ sidereal phases:}} each frequency is measured in {$N$}
intervals, each with {$1/N$} of the planned time, {$T_{\rm sid}/N$}  apart
(Sec.~\ref{sec:strategy}), with the present orientations (FLASH axis
horizontal at azimuth $57^\circ$, BabyIAXO magnet horizontal).
\textit{Best pointing:} the BabyIAXO  {bore} is tilted so that the sensitive
direction of the cavity mode is at $54.7^\circ$ from the Earth's rotation
axis; the orientation of FLASH is fixed.}
\label{tab:gain}
\end{table*}

In this work we focus on the dark photon (DP) $X_\mu$, the gauge boson of a hidden $U(1)$ that interacts with the SM through kinetic mixing with the photon~\cite{Okun:1982xi,Holdom:1985ag} (see Refs.~\cite{Fabbrichesi:2020wbt,Caputo:2021eaa} for reviews).
The Lagrangian of the minimal model is
\begin{equation}
    \mathcal{L} = -\frac{1}{4} F_{\mu\nu} F^{\mu\nu} - \frac{1}{4} X_{\mu\nu} X^{\mu\nu} + \frac{\chi}{2} F_{\mu\nu} X^{\mu\nu} + \frac{1}{2} m^2 X_\mu X^\mu\,,
\label{eq:Lagrangian}
\end{equation}
where $F_{\mu\nu}$ and $X_{\mu\nu}$ are the Maxwell field strengths of the photon and of the DP, $\chi\ll1$ is the kinetic mixing parameter, and $m$ is the DP mass, which can be generated, e.g., by the Stueckelberg mechanism~\cite{Ruegg:2003ps}.
The kinetic mixing is the only coupling of the DP to the SM, and the phenomenology is determined by the two parameters $m$ and $\chi$.
A massive DP can constitute the entire DM if produced by the misalignment mechanism~\cite{Nelson:2011sf} with non-minimal couplings to gravity~\cite{Arias:2012az,Alonso-Alvarez:2019ixv}, by inflationary fluctuations~\cite{Graham:2015rva,Kolb:2020fwh}, at the end of inflation~\cite{Bastero-Gil:2018uel}, by the decay or the oscillations of an axion or of another scalar field~\cite{Agrawal:2018vin,Co:2018lka,Dror:2018pdh}, or by a network of cosmic strings~\cite{Long:2019lwl}.
Cosmological and astrophysical observations constrain the kinetic mixing~\cite{Arias:2012az,An:2013yfc,McDermott:2019lch,Witte:2020rvb,Caputo:2020bdy}, but leave a large region of parameter space open to laboratory searches~\cite{Caputo:2021eaa}.

Unlike axions, DPs convert into photons without an external magnetic field.
From Eq.~\eqref{eq:Lagrangian}, a non-relativistic DP DM background acts on the electromagnetic field as an effective current $\mathbf{J}_{\rm eff}\propto\chi m^2\mathbf{X}$, which excites the modes of a resonant cavity whose electric field overlaps with the direction of $\mathbf{X}$~\cite{Arias:2012az,Caputo:2021eaa}.
Axion haloscopes~\cite{Sikivie:1983ip,Irastorza:2018dyq,Sikivie:2020zpn} are therefore sensitive to DP DM, and their results can be reinterpreted as limits on $\chi$~\cite{Arias:2012az,Ghosh:2021ard,Caputo:2021eaa}.

Since the DP is a vector field, the signal depends on the projection of its polarization, the unit vector $\hat{\bf X}$ along $\mathbf{X}$, on the sensitive direction of the resonant mode.
Whether the DM field keeps a fixed polarization over laboratory scales {and over the duration of a measurement campaign} depends on its production mechanism and on its evolution in the Galactic halo, and is still an open question~\cite{Caputo:2021eaa,Chen:2024vgh,Amaral:2024tjg}.
If the polarization changes randomly on time scales shorter than the measurement, the signal power of a detector sensitive to a single direction is reduced by the geometric average $\langle\cos^2\theta\rangle=1/3$.
If instead the polarization is fixed, the projection is unknown and changes with the rotation of the Earth, since the sensitive direction of a detector at rest in the laboratory precesses around the rotation axis once per sidereal day, and a robust limit must account for the probability of an unfavourable alignment~\cite{Caputo:2021eaa}.
The fixed-polarization scenario is the conservative one, and the one that sets the most demanding requirements on the measurement strategy.
Two approaches have been proposed to reduce this suppression: exploiting the rotation of the Earth through the orientation and the measurement schedule~\cite{Caputo:2021eaa}, and reading out three orthogonal degenerate modes of the cavity, whose coherent sum raises the DP form factor up to that of the axion and determines the direction of the polarization~\cite{Navarro-Madrid:2025qtm}.
The second requires a dedicated cavity design, while the first applies to any haloscope.

Two large-volume haloscopes will soon explore the $\mu$eV mass range.
{FLASH, at the INFN Laboratori Nazionali di Frascati, will search for axion DM in the frequency range 130--249\,MHz, corresponding to axion masses of approximately 0.54--1.03\,$ \mu$eV, using the primary mode of a copper cavity inside the FINUDA superconducting solenoid. A possible future upgrade would be a second cavity designed to extend the search to higher frequencies and axion masses up to approximately 1.5\,$\mu$eV~\cite{Alesini:2023qed,Calligaris:2026nno}.}
BabyIAXO, under construction at DESY as the intermediate stage of the International Axion Observatory~\cite{IAXO:2019mpb,IAXO:2020wwp,IAXO:2025ltd}, is designed as a helioscope, but its two magnet bores can host the resonant cavities developed by the RADES collaboration~\cite{Melcon:2018dba,CAST:2020rlf,Diaz-Morcillo:2021psa} for a DM search in the $1$--$2\,\mu$eV range~\cite{Ahyoune:2023gfw}.

Both programs go beyond the axion: FLASH also targets scalar DM, chameleons, DPs and high-frequency gravitational waves~\cite{Alesini:2023qed}, while the BabyIAXO infrastructure has been studied for solar DPs~\cite{OShea:2023gqn} and for high-frequency gravitational waves~\cite{Valero:2024ncz}.
{Since DPs do not require the magnetic field, the RADES cavities can search for DP DM in BabyIAXO with the magnet off and without solar tracking~\cite{Ahyoune:2023gfw}.}

Both collaborations have presented projections for DP DM, obtained by recasting their axion projections~\cite{Alesini:2023qed,Ahyoune:2023gfw}.
Ref.~\cite{Alesini:2023qed} considered two cases: $\cos^2\theta=1/3$, the average over an isotropic distribution of the DP field, and, for a fixed and unknown polarization, $\cos^2\theta=0.0025$, the value exceeded with $95\%$ probability when all directions are equally likely~\cite{Arias:2012az,Ghosh:2021ard}.
This second value results from requiring a $95\%$ probability on the polarization and, separately, a confidence level on the noise.
As pointed out in Ref.~\cite{Caputo:2021eaa}, this prescription is overly conservative: the two sources of fluctuation must be combined in a single statistical requirement.
For a single short measurement, a $95\%$ C.L.\ DP limit then corresponds to $\langle\cos^2\theta\rangle\simeq 0.024$ if the axion projection is at $95\%$ C.L., and to $\langle\cos^2\theta\rangle\simeq0.019$ if it is at $90\%$ C.L., as for FLASH~\cite{Alesini:2023qed}.
Ref.~\cite{Ahyoune:2023gfw} recast the BabyIAXO axion projection, an average $95\%$ C.L.\ upper limit, with $\langle\cos^2\theta\rangle_T\simeq0.019$, quoted from Ref.~\cite{Caputo:2021eaa} for a fixed polarization.
In that work, $\theta$ is the angle between the polarization and the vertical axis of the setup, i.e.\ the direction of the magnetic field, which for the TE$_{111}$ mode considered coincides with the sensitive direction of the mode.
In general, however, the sensitive direction is set by the electric field of the mode, not by the external magnetic field: for the orthogonal quasi-TE$_{111}$ mode of the same cavities, which can be read out with a second port~\cite{Valero:2024ncz}, it is horizontal and perpendicular to the field.
Moreover, both scan strategies measure each frequency only once, in a single step of a few minutes ($10$\,min for FLASH~\cite{Alesini:2023qed}, $9$\,min for BabyIAXO~\cite{Ahyoune:2023gfw}), with no reference to the sidereal time, and therefore do not exploit the rotation of the Earth proposed in Ref.~\cite{Caputo:2021eaa}.
Since neither experiment has started taking data, the measurement strategy can still be designed to remove these limitations.

The purpose of this paper is to define the data-taking and analysis strategy of the DP DM searches of FLASH and BabyIAXO before their first campaigns.
The theoretical ingredients are contained in Ref.~\cite{Caputo:2021eaa}, which showed that the dependence of the signal on a fixed polarization can be reduced, and in some cases removed, through the orientation of the detector and the times at which it measures.
In particular, Ref.~\cite{Caputo:2021eaa} identified the latitudes at which a North- or Zenith-pointing detector measuring for whole sidereal days is insensitive to the polarization, and showed that splitting a short measurement into three, spaced by $8$ sidereal hours, improves the sensitivity by up to an order of magnitude.
We cast these results in a form that applies directly to a frequency scan with an arbitrary orientation: measuring each frequency in $N\geq3$ sub-dwells of equal duration, equally spaced over a sidereal day, reproduces exactly the average over a full sidereal day, and if the sensitive direction of the mode makes an angle $\arccos(1/\sqrt{3})\simeq54.7^\circ$ with the Earth's rotation axis, the response is the same as for a randomly polarized DP, for any fixed polarization.
For each experiment we then specify the scan schedule, the orientation, the statistical treatment of the data, and we compute the resulting sensitivity.
The schedule requires no hardware modification, keeps the total exposure per frequency and leaves the axion search unaffected.
For BabyIAXO, whose mount is free during the magnet-off run, we also discuss how the pointing of the bore, which sets the orientation of the cavities, can further reduce the dependence on the polarization.
Additionally, we also propose to read out both TE$_{111}$ modes of the RADES cavities.
The second TE$_{111}$ mode, which has no axion coupling but a comparable DP form factor, can be read in parallel with a second port, adding exposure and a second sensitive direction.
With respect to the published projections, the expected reach in $\chi$ improves by about one order of magnitude for FLASH and by a factor of about four for BabyIAXO.

The paper is organized as follows. In Section~\ref{sec:signal} we derive the dependence of the dark-photon signal on 
a fixed polarization for an arbitrary orientation and measurement schedule, and define the statistical treatment used to recast axion limits. 
In Section~\ref{sec:experiments} we describe FLASH and the RADES cavities in BabyIAXO as dark-photon haloscopes and revisit their published projections. In Section~\ref{sec:strategy} we present the data-taking strategy, namely the sidereal-locked scan for both experiments and the pointing and mode readout for BabyIAXO, together with the resulting sensitivities. We summarize our conclusions in Sec.~\ref{sec:conclusions}. Appendix~\ref{app:direction} collects the derivation of the time-averaged response for general schedules, and Appendix~\ref{app:RADES-BabyIAXO} details the pointing optimization of the RADES-BabyIAXO cavities over their tuning range.

%=====================================================================
%=====================================================================
\section{Dark-photon signal for a fixed polarization}
\label{sec:signal}
%=====================================================================
%=====================================================================

In this section we review the DP signal in a resonant cavity and its dependence on the sensitive direction of the mode, give the time-averaged response to a fixed polarization for any orientation and measurement schedule, and define the conversion factors that replace the random-polarization value $1/3$ in the recast of axion limits.
The theoretical ingredients are in Ref.~\cite{Caputo:2021eaa}, whose notation we follow; we write them in a form that applies directly to a frequency scan, and collect the derivations in Appendix~\ref{app:direction}.

For non-relativistic DM, $X^\mu\simeq(X^0,\mathbf{X})\,e^{-imt}$, with $X^0\simeq\mathbf{v}\cdot\mathbf{X}$ suppressed by the DM velocity $v\sim10^{-3}$.
To leading order in $\chi$, Eq.~\eqref{eq:Lagrangian} adds to Maxwell's equations the effective current $\mathbf{J}_{\rm eff}=\chi m^2\mathbf{X}$, which in free space generates an electric field along the polarization $\hat{\bf X}$ with amplitude $\chi\sqrt{2\rho_{\rm DM}}$, where $\rho_{\rm DM}$ is the local DM density~\cite{Caputo:2021eaa}.
The field is coherent over the length $\lambda_c=2\pi/(mv)\simeq1.2\,{\rm km}\,(1\,\mu{\rm eV}/m)$ and the time $\tau_c=2\pi/(mv^2)\simeq4\,{\rm ms}\,(1\,\mu{\rm eV}/m)$, and is therefore uniform over the detector.
On resonance, for a loaded quality factor $Q_L$ below the DM quality factor $\sim10^6$, the power extracted from a cavity mode with electric field $\mathbf{E}(\mathbf{x})$ is~\cite{Arias:2012az,Caputo:2021eaa}
\begin{equation}
P_X=\kappa\,\chi^2 m\,\rho_{\rm DM}\,V\,Q_L\,C_X\,,
\label{eq:PX}
\end{equation}
where $V$ is the volume of the cavity, $\kappa=\beta/(1+\beta)$, with $\beta$ the coupling of the readout port, and $C_X$ is the DP form factor.
{
For a standing-wave mode, the electric field $\mathbf{E}$ is real in space (up to a global sinusoidal time oscillation and small corrections from cavity losses and ports), 
and $\int dV\,\mathbf{E}$ defines a vector $\hat {\bf n}$ pointing in a fixed direction, the polarization-sensitive direction 
of the mode, namely 
$\hat{{\bf n}} = \int dV\,\mathbf{E}/|\int dV\,\mathbf{E}|$.}
% The form factor then factorizes,
{Since $\hat{\bf X}$ is uniform over the cavity, the form factor factorizes,}
\begin{equation}
C_X=\frac{\left|\int dV\,\mathbf{E}\cdot\hat{\bf X}\right|^2}{V\int dV\,|\mathbf{E}|^2}=C^{(0)}\cos^2\theta\,,
\label{eq:CX}
\end{equation}
with $\cos\theta=\hat{\mathbf{n}}\cdot\hat{\bf X}$ and $C^{(0)}=|\int dV\,\mathbf{E}|^2/(V\int dV\,|\mathbf{E}|^2)$ being the form factor for $\hat{\bf X}\parallel\hat{\mathbf{n}}$.
{Both $\hat{\mathbf{n}}$ and $C^{(0)}$ are properties of the mode alone, independent of the magnetic field and of the readout port, unless other modes are degenerate with it within the linewidth~\cite{Navarro-Madrid:2025qtm}.}

For the $\mathrm{TM}_{010}$ mode of a cylindrical cavity in a solenoid with magnetic field $\mathbf{B}_0$, as in FLASH, $\hat{\mathbf{n}}$ is the cavity axis parallel to $\mathbf{B}_0$, and, for an ideal cylinder, $C^{(0)} = 4/x_{01}^2 \simeq 0.69$, equal to the axion form factor.\footnote{{Here, $x_{01} \approx 2.405$ denotes the first positive root of the Bessel function $J_0(x)$ (satisfying $J_0(x_{01}) = 0$), while $x'_{11} \approx 1.841$ denotes the first positive root of the derivative $J_1'(x)$ (satisfying $J_1'(x'_{11}) = 0$), which governs transverse-electric (TE) or alternative mode configurations.}}
The RADES cavities in the BabyIAXO bores have two TE$_{111}$ modes, split in frequency by the shape of the cross-section and by the tuning plates, with $\hat{\mathbf{n}}$ transverse to the bore: along $\mathbf{B}_0$ for one mode (Mode~1) and perpendicular to it for the other (Mode~2; Sec.~\ref{sec:experiments}), with $C^{(0)}=16/[\pi^2(x_{11}'^2-1)]\simeq0.68$ for an ideal cylinder.
Tuning elements can rotate $\hat{\mathbf{n}}$ if they break the symmetry of the cavity ({Sec.~\ref{sec:experiments}}). 

The same data constrain the axion-photon coupling and the kinetic mixing.
An axion search in a uniform field $B_0$ measures the power $P_a=\kappa\,\gagg^2B_0^2\,(\rho_{\rm DM}/m)\,V\,Q_L\,C_a$, with $C_a$ the axion form factor, so that a limit on {the axion-photon coupling} $\gagg$ translates into
\begin{equation}
\chi=\frac{\gagg B_0}{m}\left(\frac{C_a}{C^{(0)}\,\langle\cos^2\theta\rangle^{\rm conv}}\right)^{1/2},
\label{eq:recast}
\end{equation}
where the conversion factor $\langle\cos^2\theta\rangle^{\rm conv}$, defined in Eq.~\eqref{eq:conv} below, accounts for the unknown polarization, the measurement schedule and the statistics.
We emphasize that in previous literature—frequently within the context of axion–dark photon recasts—$\theta$ has often been defined as the geometric angle between $\mathbf{X}$ and the magnetic field $\mathbf{B}_0$, yielding the form-factor relation $C_X = C_a\cos^2\theta$~\cite{Arias:2012az,Ghosh:2021ard,Caputo:2021eaa}. Strictly speaking, however, this direct correspondence holds only when $\hat{\mathbf{n}}\parallel\mathbf{B}_0$ and $C_a = C^{(0)}$. {For a mode with $\hat{\mathbf{n}}\perp\mathbf{B}_0$, such as Mode~2 in RADES cavities, $C_a \simeq 0$; in this case, the axion limit cannot be recovered via recasting, and the DP sensitivity must be computed directly from Eq.~\eqref{eq:PX}, with $\theta$ measured relative to $\hat{\mathbf{n}}$.}
The polarization can change on the time scale $\tau_c$ or remain fixed over a campaign~\cite{Caputo:2021eaa,Chen:2024vgh,Amaral:2024tjg}.
In the first case (random polarization), a measurement much longer than $\tau_c$ averages $\cos^2\theta$ over all directions and gives $1/3$, independently of the orientation and of the time of the measurement.
In the second case (fixed polarization), $\hat{\bf X}$ is a single unknown direction, which we describe with an isotropic prior, and $\cos\theta$ changes with the rotation of the Earth.
We take the fixed polarization as the baseline for the strategy and the random one as the reference sensitivity.

\begin{figure}[t]
\centering
\includegraphics[width=0.8\columnwidth]{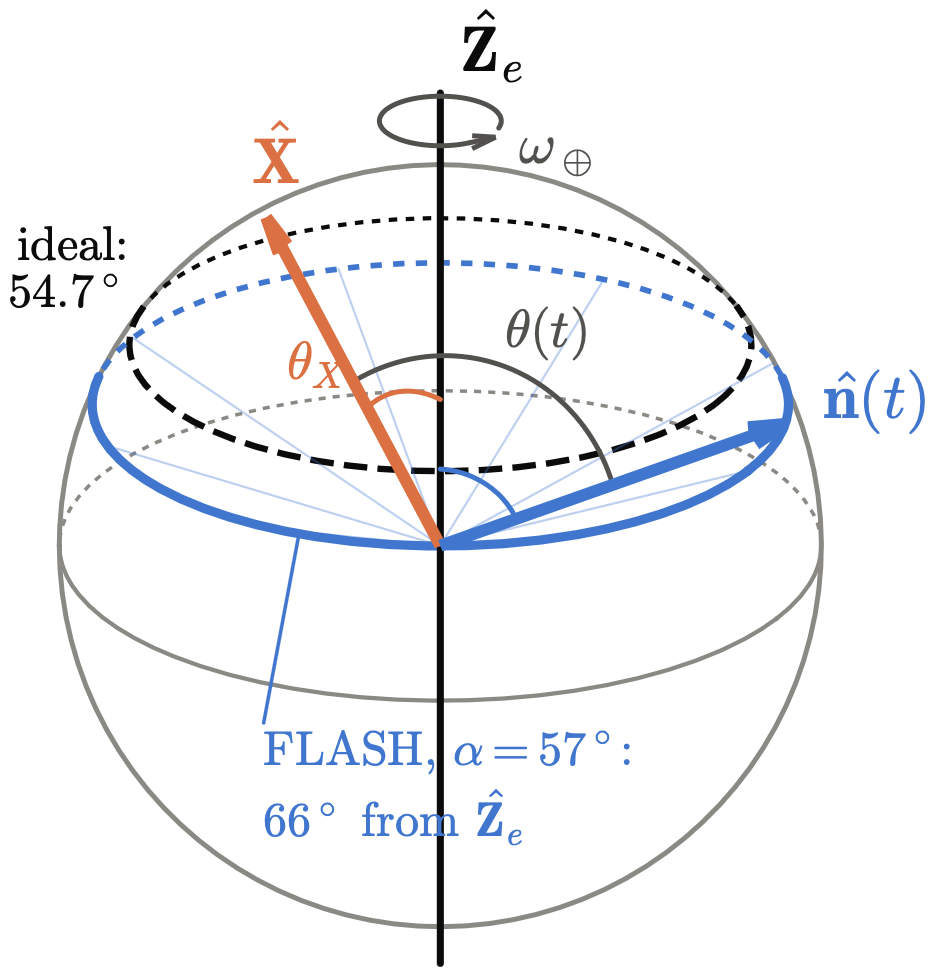}
\caption{Polarization $\hat{\bf X}$ and sensitive direction $\hat{\mathbf{n}}(t)$ of a detector at rest in the laboratory, in the equatorial frame ($\hat{\bf Z}_e$ along the rotation axis of the Earth).
$\hat{\bf X}$ is fixed, at polar angle $\theta_X$, while $\hat{\mathbf{n}}$ precesses once per sidereal day on a cone around $\hat{\bf Z}_e$, here at $66^\circ$  for the FLASH axis (horizontal, azimuth $57^\circ$); $\theta(t)$ is the angle between the two.
If $\hat{\mathbf{n}}$ lies on the cone at $54.7^\circ$ (dashed), the signal averaged over the sidereal day does not depend on the polarization (Sec.~\ref{sec:signal}).}
\label{fig:geometry}
\end{figure}

We describe $\hat{\bf X}$ in the equatorial frame of Fig.~\ref{fig:geometry} by its polar angle $\theta_X$ and azimuth $\phi_X$, both constant.
The sensitive direction is fixed in the laboratory, with azimuth $\alpha$ from North to West (only $\cos\alpha$ enters) and elevation $\beta$ above the horizon, at latitude $\lambda$.
Its projection on the rotation axis $\hat{\bf Z}_e$,
\begin{equation}
n_z=\hat{\mathbf{n}}\cdot\hat{\bf Z}_e=\cos\beta\cos\alpha\cos\lambda+\sin\beta\sin\lambda\,,
\label{eq:nz}
\end{equation}
is constant, while its component orthogonal to $\hat{\bf Z}_e$ rotates with the sidereal angular frequency $\omega_\oplus=2\pi/T_{\rm sid}$, with $T_{\rm sid}=23$\,h\,$56$\,min\,$4$\,s. 
The projection on the polarization is then~\cite{Caputo:2021eaa}
\begin{equation}
\cos\theta(t)=\sqrt{1-n_z^2}\,\sin\theta_X\cos(\omega_\oplus t-\psi)+n_z\cos\theta_X\,,
\label{eq:costheta_t}
\end{equation}
where the phase $\psi$ depends on $\phi_X$, on the orientation and on the origin of time.
Squaring Eq.~\eqref{eq:costheta_t},
\begin{equation}
\begin{split}
\cos^2\theta(t)={}&n_z^2\cos^2\theta_X+\frac{1-n_z^2}{2}\sin^2\theta_X\\
+&2n_z\sqrt{1-n_z^2}\,\sin\theta_X\cos\theta_X\cos(\omega_\oplus t-\psi)\\
+&\frac{1-n_z^2}{2}\sin^2\theta_X\cos(2\omega_\oplus t-2\psi)\,,
\end{split}
\label{eq:harmonics}
\end{equation}
a constant plus a first and a second harmonic of the sidereal frequency, with periods $T_{\rm sid}$ and $T_{\rm sid}/2$.

If the exposure of a frequency is split into sub-dwells, the optimally combined excess power is proportional to the average of $\cos^2\theta(t)$ over the sub-dwells, weighted by their exposure for equal noise, which we denote $\langle\cos^2\theta\rangle$.
Since the time dependence of $\cos^2\theta(t)$ is only in the two harmonics of Eq.~\eqref{eq:harmonics}, $\langle\cos^2\theta\rangle$ depends in general on $\theta_X$, $\phi_X$ and the schedule.
Both harmonics average to zero for an exposure of an integer number of sidereal days~\cite{Caputo:2021eaa} and, more usefully for a frequency scan, for $N\geq3$ sub-dwells of equal duration and noise with start times spaced by $T_{\rm sid}/N$, whatever their duration {$T$. For this sampling schedule the average is defined as
\begin{equation}
    \langle \cos^2\theta\rangle = \frac{1}{NT} \sum_{k=0}^{N-1} \int_{t_0+kT_{\rm sid}/N}^{t_0+kT_{\rm sid}/N+T} \cos^2\theta(t)\,dt\,,
\label{eq:avg}
\end{equation}
being $t_0$ the starting time of the first measurement
} (see Appendix~\ref{app:direction} for more details).

In both cases, which we call full sidereal coverage,
\begin{equation}
\begin{split}
\langle\cos^2\theta\rangle&=\frac{1-n_z^2}{2}+\frac{3n_z^2-1}{2}\cos^2\theta_X\\
&=\frac13+\frac23\,P_2(n_z)\,P_2(\cos\theta_X)\,,
\end{split}
\label{eq:fullcov}
\end{equation}
with $P_2(x)=(3x^2-1)/2$.
The condition $N\geq3$ is necessary, since two sub-dwells spaced by $T_{\rm sid}/2$ cancel the first harmonic but not the second.
The case $N=3$ corresponds to the three measurements spaced by $8$ sidereal hours considered in Ref.~\cite{Caputo:2021eaa}.
For $n_z=0$ the first harmonic is absent, and two sub-dwells spaced by $T_{\rm sid}/4$ suffice, in agreement with the $6$\,h spacing found in Ref.~\cite{Caputo:2021eaa} for West-pointing detectors.
For other schedules $\langle\cos^2\theta\rangle$ depends also on $\phi_X$ (Appendix~\ref{app:direction}), which we use in Sec.~\ref{sec:strategy} to quantify the effect of deviations from the planned schedule.

Equation~\eqref{eq:fullcov} has three consequences.
First, the site and the orientation enter only through $n_z^2$.
Second, for an isotropic prior $|\cos\theta_X|$ is uniformly distributed in $[0,1]$, and the response has mean $1/3$ for any orientation, but a spread that depends on it: it ranges between $n_z^2$, for $\hat{\bf X}$ along the rotation axis, and $(1-n_z^2)/2$, for $\hat{\bf X}$ in the equatorial plane, with variance $(3n_z^2-1)^2/45$.
Third, the spread vanishes for $P_2(n_z)=0$, i.e.\ when $\hat{\mathbf{n}}$ makes the angle $\arccos(1/\sqrt{3})\simeq54.7^\circ$ with the rotation axis of the Earth 
\begin{equation}
\cos\beta\cos\alpha\cos\lambda+\sin\beta\sin\lambda=\pm\frac{1}{\sqrt{3}}\,.
\label{eq:ideal}
\end{equation}
Then $\langle\cos^2\theta\rangle=1/3$ for every polarization, as for a random one.
The latitudes found in Ref.~\cite{Caputo:2021eaa}, $\pm54.7^\circ$ for a North-pointing direction and $\pm35.3^\circ$ for a Zenith-pointing one, are the special cases $\alpha=\beta=0$ and $\beta=90^\circ$ of Eq.~\eqref{eq:ideal}.
A horizontal $\hat{\mathbf{n}}$ is ideal for $\cos\alpha=1/(\sqrt{3}\cos\lambda)$, which has a solution for $|\lambda|\leq54.7^\circ$ ($\alpha=39.2^\circ$ at FLASH and $13.5^\circ$ at BabyIAXO), while an East--West direction ($n_z=0$) never is.
In the opposite limit, $n_z^2=1$, the response is $\cos^2\theta_X$, which has the same distribution as the response to a single exposure much shorter than a sidereal day, at any orientation.

\begin{figure}[t]
\centering
\includegraphics[width=\columnwidth]{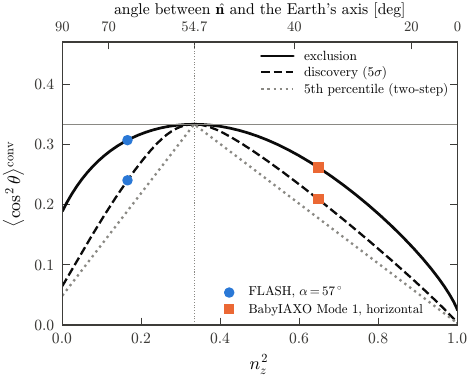}
\caption{Conversion factors at $95\%$ C.L.\ for full sidereal coverage, Eqs.~\eqref{eq:master} and~\eqref{eq:conv}, for exclusion (solid) and $5\sigma$ discovery (dashed), versus $n_z^2$ (bottom axis) or the angle between $\hat{\mathbf{n}}$ and the rotation axis of the Earth (top axis).
Dotted: 5th percentile of $\langle\cos^2\theta\rangle$, the factor of the two-step prescription discussed in the text.
The horizontal line is the random-polarization value $1/3$.
Markers: FLASH with a horizontal axis at azimuth $57^\circ$ (circles); BabyIAXO, Mode~1 with the magnet horizontal (squares), as in Ref.~\cite{Ahyoune:2023gfw}. }
\label{fig:conv}
\end{figure}

Following Ref.~\cite{Caputo:2021eaa}, we model the excess power measured at a given frequency, after combining the sub-dwells, as $P=P_X'\langle\cos^2\theta\rangle+N$, where $P_X'\propto\chi^2$ is the power for $\hat{\bf X}\parallel\hat{\mathbf{n}}$ and $N$ is a Gaussian noise with standard deviation $\sigma_N$.
We quote median expected sensitivities, i.e.\ we set the observed excess to zero.
An axion limit at confidence level CL corresponds to the signal power $P_a=z_{\rm CL}\,\sigma_N$, with $z_{\rm CL}=\Phi^{-1}({\rm CL})$ and $\Phi$ the cumulative distribution of a standard Gaussian ($z_{\rm CL}=1.645$ at $95\%$ C.L., one-sided).
For a fixed polarization, the probability of a downward fluctuation must include both the noise and the polarization.
With full sidereal coverage and $c\equiv|\cos\theta_X|$ uniformly distributed in $[0,1]$, the condition reads~\cite{Caputo:2021eaa}
\begin{equation}
\int_0^1 dc\;\Phi\!\left(z_{\rm obs}-\frac{P_X'}{\sigma_N}\left[\frac{1-n_z^2}{2}+\frac{3n_z^2-1}{2}\,c^2\right]\right)=1-{\rm CL}\,,
\label{eq:master}
\end{equation}
with $z_{\rm obs}=0$ for an exclusion limit and $z_{\rm obs}=5$ for a discovery, i.e.\ for a signal that exceeds $5\sigma$ with probability CL~\cite{Caputo:2021eaa}.
The solution for $P_X'/\sigma_N$ defines the conversion factor
\begin{equation}
\langle\cos^2\theta\rangle^{\rm conv}=\frac{(z_{\rm obs}+z_{\rm CL})\,\sigma_N}{P_X'}\,,
\label{eq:conv}
\end{equation}
denoted $\langle\cos^2\theta\rangle^{\rm excl}$ and $\langle\cos^2\theta\rangle^{\rm disc}$ in the two cases: the known projection factor that would give the same sensitivity.
It replaces $1/3$ in Eq.~\eqref{eq:recast}, so that $\chi_{\rm lim}\propto(\langle\cos^2\theta\rangle^{\rm conv})^{-1/2}$, and depends only on $n_z^2$ and on the CL (Fig.~\ref{fig:conv} and Table~\ref{tab:conv}).
For $n_z^2=1/3$ it equals $1/3$, for exclusion and discovery and at any CL, and for any other orientation it is smaller (Sec.~\ref{sec:strategy_flash}).
We quantify the loss with
\begin{equation}
F=\left(\frac{1/3}{\langle\cos^2\theta\rangle^{\rm conv}}\right)^{1/2},
\label{eq:F}
\end{equation}
the ratio of the limit on $\chi$ to the one for the ideal orientation, which is also the one for a random polarization.
The loss is $F^2$ in signal power and, for a radiometer-limited search, $F^4$ in integration time.
Near the ideal orientation the loss is quadratic, $F_{\rm excl}\simeq1+0.135\,(3n_z^2-1)^2$: an angle within about $12^\circ$ of the ideal one costs less than $5\%$ in $\chi$ for exclusion, and within about $6^\circ$ for discovery.

\begin{table}[t]
\caption{Conversion factors at $95\%$ C.L.\ for full sidereal coverage, as functions of $n_z^2$ (any site and orientation), and loss factors $F$ in $\chi$ with respect to the ideal orientation, Eq.~\eqref{eq:F}.
The second column is the angle between $\hat{\mathbf{n}}$ and the rotation axis of the Earth $\hat{\bf Z}_e$.
The last row, $n_z^2=1$, also gives the factors for a single exposure much shorter than a sidereal day, at any orientation.\vspace{0.15cm}}
\label{tab:conv}
\begin{ruledtabular}
\begin{tabular}{cccccc}
$n_z^2$ & angle & $\langle\cos^2\theta\rangle^{\rm excl}$ & $\langle\cos^2\theta\rangle^{\rm disc}$ & $F_{\rm excl}$ & $F_{\rm disc}$ \\
\hline
$0$ & $90^\circ$ & $0.189$ & $0.065$ & $1.328$ & $2.265$ \\
$0.1$ & $71.6^\circ$ & $0.278$ & $0.176$ & $1.095$ & $1.376$ \\
$0.2$ & $63.4^\circ$ & $0.317$ & $0.271$ & $1.025$ & $1.109$ \\
$1/3$ & $54.7^\circ$ & $1/3$ & $1/3$ & $1$ & $1$ \\
$0.5$ & $45.0^\circ$ & $0.312$ & $0.281$ & $1.034$ & $1.089$ \\
$2/3$ & $35.3^\circ$ & $0.253$ & $0.199$ & $1.148$ & $1.294$ \\
$1$ & $0^\circ$ & $0.024$ & $0.0034$ & $3.727$ & $9.901$ \\
\end{tabular}
\end{ruledtabular}
\end{table}

The last row of Table~\ref{tab:conv} also gives the factors for a single exposure much shorter than a sidereal day, at any orientation: $\langle\cos^2\theta\rangle^{\rm excl}=0.024$ and $\langle\cos^2\theta\rangle^{\rm disc}=0.0034$ ($0.025$ and $0.0036$ in Ref.~\cite{Caputo:2021eaa}).
With respect to full coverage at the ideal orientation, this costs a factor $3.7$ in $\chi$ for exclusion and $10$ for discovery.
This is the loss of the scan strategies of Refs.~\cite{Alesini:2023qed,Ahyoune:2023gfw}, which the schedule of Sec.~\ref{sec:strategy} removes.

Two prescriptions found in the literature give different factors.
Replacing the unknown $\langle\cos^2\theta\rangle$ by its 5th percentile and then applying the axion criterion imposes the $95\%$ requirement twice, once on the polarization and once on the noise~\cite{Caputo:2021eaa}.
It gives the dotted line in Fig.~\ref{fig:conv}: $0.0025$ instead of $0.024$ for a single short exposure, and $0.049$ instead of $0.189$ for an East--West direction with full coverage.
The CL enters in two places, the right-hand side of Eq.~\eqref{eq:master} and $z_{\rm CL}$ in Eq.~\eqref{eq:conv}, and changing it only in the second converts an axion limit at one CL into a DP limit at another.
The factor $0.019=0.024\times z_{90}/z_{95}$ of Ref.~\cite{Caputo:2021eaa} is a $95\%$ C.L.\ DP limit obtained from a $90\%$ C.L.\ axion limit; with both at $90\%$, the same exposure gives $0.076$.
Our DP limits are at $95\%$ C.L.
For BabyIAXO the axion projection is also at $95\%$ C.L.~\cite{Ahyoune:2023gfw}, so that Table~\ref{tab:conv} applies directly, and the factor $0.019$ used in Ref.~\cite{Ahyoune:2023gfw} is conservative by $(z_{95}/z_{90})^{1/2}=1.13$ in $\chi$.
For FLASH, whose axion projection is at $90\%$ C.L.~\cite{Alesini:2023qed}, we quote $95\%$ C.L.\ DP limits from it, i.e.\ we multiply the factors of Table~\ref{tab:conv} by $z_{90}/z_{95}=0.78$, which leaves $F$ unchanged.

%=====================================================================
\section{FLASH and BabyIAXO as dark-photon haloscopes}
\label{sec:experiments}

\begin{figure*}[t]
\centering
\includegraphics[width=0.98\textwidth]{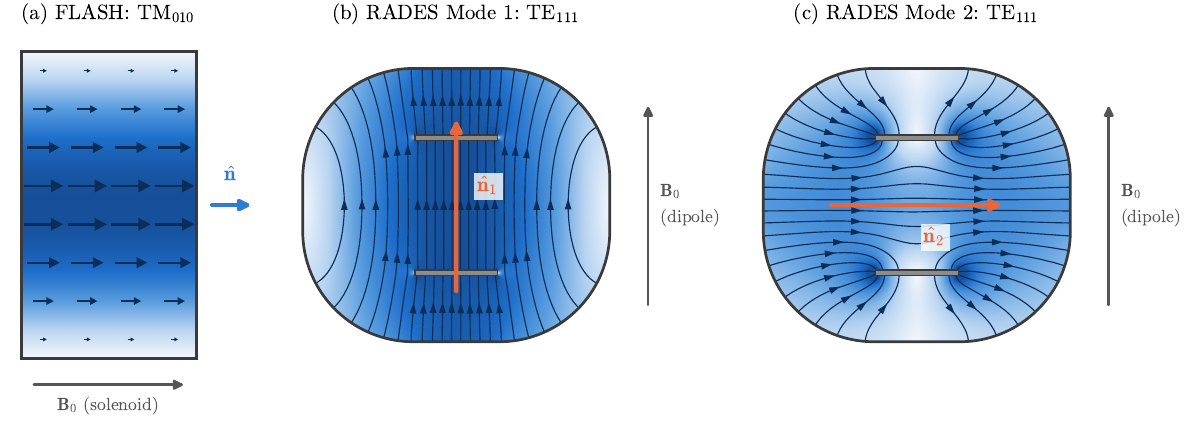}
\caption{Electric field of the modes considered here and their sensitive directions $\hat{\mathbf{n}}$, the directions of $\int dV\,\mathbf{E}$ (thick arrows).
(a)~TM$_{010}$ mode of FLASH, side view{, for an ideal cylinder}: $\hat{\mathbf{n}}$ is along the axis of the cavity and of the solenoid.
(b, c)~The two TE$_{111}$ modes of a RADES cavity, cross-section at mid-length{, computed for the first RADES cavity ($56\times50.4$\,cm) with the two tuning plates (gray) at $0^\circ$}: the field lines curve, but $\hat{\mathbf{n}}$ points along the height (Mode~1) or the width (Mode~2) of the cross-section.
{The static field $\mathbf{B}_0$ of the solenoid is shown in (a) for reference; it does not set $\hat{\mathbf{n}}$.
In the BabyIAXO dipole, $\mathbf{B}_0$ is along the height of the cross-section, so that only Mode~1 couples to the axion.}
{Since the cross-section is not circular, the two modes have different frequencies}.}
\label{fig:modes}
\end{figure*}

FLASH and the RADES haloscope in BabyIAXO realize the two generic configurations of cavity haloscopes: a TM$_{010}$ mode in a solenoid, with $\hat{\mathbf{n}}$ along the axis of the cavity and of the magnet, and TE$_{111}$ modes in a dipole, with $\hat{\mathbf{n}}$ transverse to the bore (Fig.~\ref{fig:modes}).
Their parameters are summarized in Table~\ref{tab:setups}.

\begin{table}[t]
\caption{Parameters of FLASH~\cite{Alesini:2023qed} and the RADES haloscope in BabyIAXO~\cite{Ahyoune:2023gfw,IAXO:2020wwp} relevant for the DP search, as assumed in the published projections.
Masses are converted from the frequency bands, and $\hat{\mathbf{n}}$ is the sensitive direction of Sec.~\ref{sec:signal}.
For BabyIAXO the cavity entry refers to the four RADES cavities, and the field is the value used for the axion projection; the DP search is planned with the magnet off. \vspace{0.15cm}}
\label{tab:setups}
\setlength{\tabcolsep}{2.0pt} % Increase this value for more space
\begin{ruledtabular}
\begin{tabular}{lll}
 & FLASH
 & BabyIAXO \\
\hline
latitude & $41.82^\circ$ & $53.58^\circ$ \\
magnet & solenoid, $1.1$\,T & dipole, $2$\,T \\
cavities [m] & {$L=1.5$\,,\,$R=0.95$} & $L=5$ \\
mode & TM$_{010}$ & TE$_{111}$ \\
band [MHz] & {$130$--$249$} & $252.8$--$469.3$ \\
mass [$\mu$eV] & {$0.54$--$1.03$} & $1.05$--$1.94$ \\
$\hat{\mathbf{n}}$ & solenoid axis & $\perp$ bore \\
orientation of $\hat{\mathbf{n}}$ & horiz., azimuth $57^\circ$ & vert. (Mode~1) \\
$n_z^2$ & $0.165$ & $0.648$ \\
dwell per step & $10$\,min & $9$\,min \\
steps & $\sim1.2\times10^5$ & $\sim7\times10^4$ \\
axion C.L. & $90\%$ & $95\%$ \\
published DP factor & $0.0025$--$1/3$ & $0.019$ \\
\end{tabular}
\end{ruledtabular}
\end{table}

%%%%%%%%%%%%%%%%%%%%%%%
\subsection{FLASH}
\label{app:flash}
%%%%%%%%%%%%%%%%%%%%%%%%

\begin{figure}[t]
    \centering
    \includegraphics[width=0.85\linewidth]{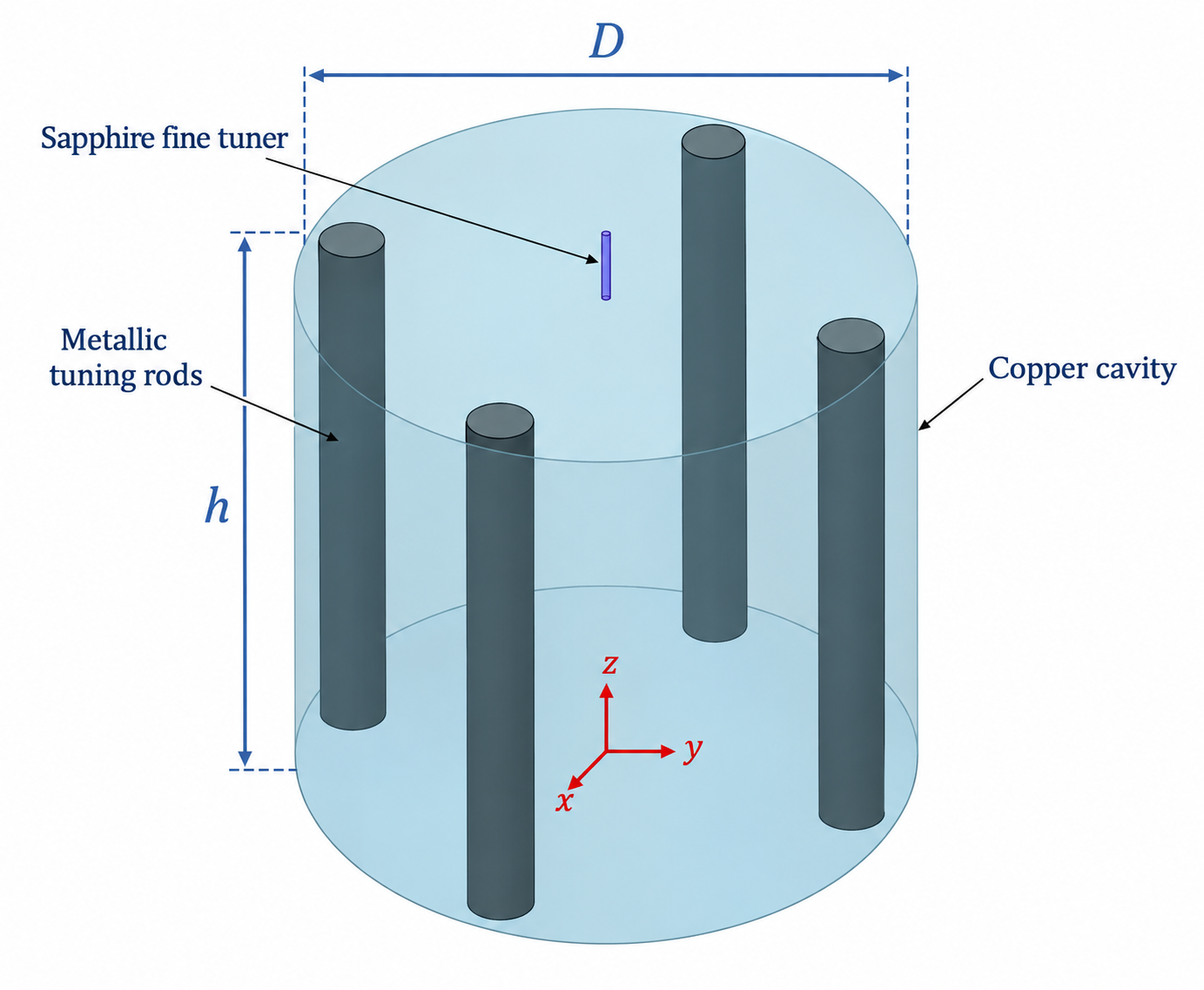}
    \caption{Schematic representation of the FLASH cylindrical
    copper cavity in its reference configuration.
    The four metallic tuning rods and the central sapphire
    fine tuner are shown in dark gray and violet, respectively.
    The cavity diameter ($D$) and length ($h$) are indicated,
    together with the Cartesian reference frame.}
    \label{fig:FLASH_geometry}
\end{figure}

FLASH will use the superconducting solenoid of the FINUDA experiment at the Laboratori Nazionali di Frascati ($41^\circ49'26''$\,N), providing an axial magnetic field of $1.1$\,T in a bore of $2.7$\,m diameter and $2.4$\,m length~\cite{Alesini:2023qed}.
The current design features a single cylindrical cavity made of oxygen-free high-conductivity copper, with a nominal internal radius of $0.95$\,m and an internal length of $1.50$\,m. Four movable metallic rods provide frequency tuning, while an axial sapphire rod provides fine frequency adjustment. The axion search will use the TM$_{010}$ mode as the primary mode over approximately $130$--$249$\,MHz with a planned integration time of $10$\,min per frequency step.
The cavity geometry and its tuning elements are schematically illustrated in Fig.~\ref{fig:FLASH_geometry}.

For the TM$_{010}$ mode, the polarization-sensitive direction is taken to coincide with the horizontal cavity and solenoid axis. A preliminary measurement gives an azimuth of $57^\circ$ from North, refining the approximate East-Northeast orientation initially reported in Ref.~\cite{Alesini:2023qed}.
For comparison with the published projection, the joint criterion of Sec.~\ref{sec:signal} replaces the conversion factor $0.0025$ used in Ref.~\cite{Alesini:2023qed} by approximately $0.019$ when deriving a $95\%$ C.L. dark-photon sensitivity from the published $90\%$ C.L. axion projection in the single-short-exposure approximation. This improves the reach in $\chi$ by a factor of approximately $2.8$ before any change to the measurement schedule.

 \begin{figure}[t]
     \centering
     \includegraphics[width=0.85\linewidth]{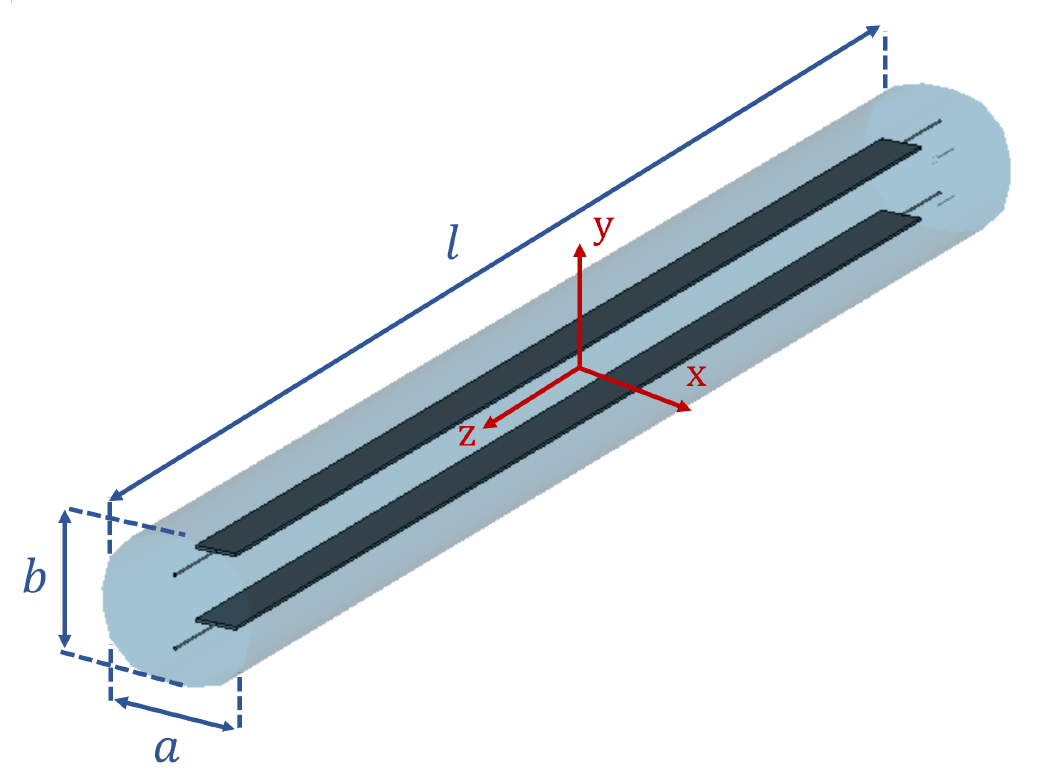}
     \caption{Schematic of the RADES-BabyIAXO cavities geometry (adapted from~\cite{Valero:2024ncz}). The cavity has a quasi-cylindrical shape, with dimensions $a$ (width), $b$ (height) and $l$ (length), where the relation $b = 0.9a$ is fulfilled. The metallic rectangular plates for the frequency tuning are shown in dark gray color. In red, the chosen reference frame in the center of the cavity~\cite{Valero:2024ncz,Ahyoune:2023gfw}.}
     \label{fig:RADES-BabyIAXO}
 \end{figure}

%%%%%%%%%%%%%%%%%%
%%%%%%%%%%%%%%%%%%
\subsection{BabyIAXO}
%%%%%%%%%%%%%%%%%%
%%%%%%%%%%%%%%%%%%

BabyIAXO, under construction at DESY in Hamburg ($53.58^\circ$\,N), is a superconducting magnet with two bores of $70$\,cm diameter and $10$\,m length that generate a predominantly dipolar field~\cite{IAXO:2020wwp,Ahyoune:2023gfw}.
It is mounted on a drive with an elevation range of $\pm25^\circ$ and full azimuthal rotation~\cite{IAXO:2020wwp}.
The RADES proposal~\cite{Ahyoune:2023gfw} foresees $5$\,m-long quasi-cylindrical cavities of four different sizes, with cross-sections ranging from $56\times50.4$\,cm to $35.5\times32.0$\,cm (see Fig.~\ref{fig:RADES-BabyIAXO}).
Tuned by two rotating plates~\cite{Ahyoune:2023gfw,Valero:2024ncz}, the four cavities collectively cover $252.8$--$469.3$\,MHz ($1.05$--$1.94\,\mu$eV); they are inserted two at a time into one of the bores and cooled to about $4.6$\,K~\cite{Ahyoune:2023gfw}.

The baseline axion projection—an average $95\%$ C.L.\ upper limit for $440$ days of effective exposure across the four cavities—corresponds to roughly $7\times10^4$ tuning steps of $9$\,min each, with BabyIAXO stationary in its parking position~\cite{Ahyoune:2023gfw}.
{Ref.~\cite{Ahyoune:2023gfw} recasts this sensitivity into a DP projection using $\langle\cos^2\theta\rangle_T\simeq0.019$, assuming Mode~1 and a horizontal magnet.}
Because the DP search requires no magnetic field, it can run with the magnet off and without solar tracking~\cite{Ahyoune:2023gfw}.
{The schedule of this run is not fixed yet; here we assume that during the run the bores can be pointed within the range of the drive.}
Consequently, the {pointing of the bores}—which fixes the orientation of the cavities—can be optimized specifically for the DP search and held fixed throughout the run.

{The RADES cavities support two TE$_{111}$-like modes with sensitive directions aligned along the height (Mode~1) and width (Mode~2) of the cross-section (Fig.~\ref{fig:modes}).
Because the cross-section is elliptical rather than circular ($b\simeq0.9\,a$~\cite{Valero:2024ncz}), this degeneracy is lifted: rotating the plates tunes Mode~1 downward and Mode~2 upward~\cite{Valero:2024ncz}, covering approximately $253$--$300$\,MHz and $303$--$333$\,MHz, respectively, in the largest cavity~\cite{Ahyoune:2023gfw,Valero:2024ncz}.
Mode~1 is used for the axion search and the published projections; for a horizontal magnet {with the tuning plates at $0^\circ$,} $\hat{\mathbf{n}}_1$ is vertical, yielding $n_z^2=\sin^2\lambda=0.648$ at DESY regardless of azimuth.
Mode~2 has its sensitive axis $\hat{\mathbf{n}}_2$ orthogonal to {that of Mode~1} and can be read out simultaneously via a second loop, rotated by $90^\circ$, at the opposite end of the cavity~\cite{Valero:2024ncz}.
During the magnet-off run, Mode~2 acts as a parallel DP channel; wherever its frequency band overlaps with Mode~1 of a smaller cavity, the same mass is probed along two orthogonal directions.
When the two plates are kept parallel as they rotate, as in Refs.~\cite{Ahyoune:2023gfw,Valero:2024ncz}, the cavity loses its mirror symmetries at intermediate plate angles {and a dedicated study of the mode behavior is required. This analysis is presented in Appendix~\ref{app:RADES-BabyIAXO}. As shown in our simulations of the first cavity in Fig.~\ref{fig:RADES modes}, both sensitive directions rotate in the transverse plane by up to about $30^\circ$.} Conversely, if the plates are rotated by opposite angles, the directions of the modes remain fixed. {This counter-rotating plate scheme is supported by measurements performed with an existing prototype for the RADES-DarkQuantum haloscope~\cite{Puente:2026mog}.}

%%%%%%%%%%%%%%%%%%
%=====================================================================
\section{Data-taking strategy}
\label{sec:strategy}
%%%%%%%%%%%%%%%%%%

\begin{figure*}[t]
\centering
\includegraphics[width=0.85\textwidth]{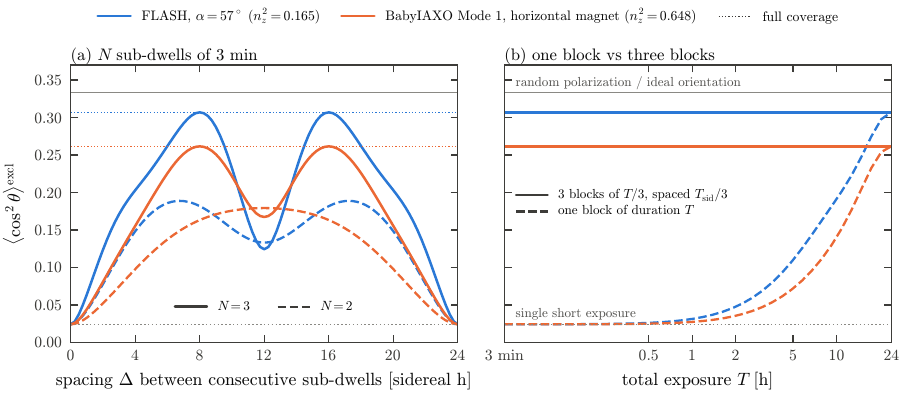}
\caption{
Exclusion conversion factor at $95\%$ C.L.\ for FLASH (axis at azimuth $57^\circ$) and BabyIAXO (Mode~1, horizontal magnet).
(a)~$N$ sub-dwells of $3$\,min, with consecutive start times spaced by $\Delta$: three sub-dwells reach full coverage (dotted lines) at $\Delta=T_{\rm sid}/3$ and $2T_{\rm sid}/3$, two never do.
(b)~The same total exposure $T$, taken as one continuous block (dashed) or as three blocks of $T/3$ spaced by $T_{\rm sid}/3$ (solid).
Gray lines: random polarization and single short exposure.}
\label{fig:schedule}
\end{figure*}
The strategy has three elements: the measurement schedule, which applies to both experiments, together with the pointing of the magnet and the choice of the modes read out, which apply to BabyIAXO.
None of them changes the cavities or the total exposure per frequency.

\subsection*{Sidereal-locked scan}
\label{sec:schedule}

Each frequency must be measured in $N\geq3$ sub-dwells of equal exposure, with start times spaced by $T_{\rm sid}/N$ (Sec.~\ref{sec:signal}).
Only the sidereal phase of each sub-dwell matters, not the day on which it is taken.
For $N=3$, a simple implementation splits the band into segments of $K$ consecutive tuning steps and scans each segment three times, with one third of the planned dwell $\tau$ per step.
With $K=T_{\rm sid}/\tau$ each sweep lasts $T_{\rm sid}/3=7$\,h\,$58.7$\,min, so that every frequency receives three sub-dwells of duration $\tau/3$, spaced by exactly $T_{\rm sid}/3$, and the same total exposure $\tau$ as in the published scans.
{
This gives $K=144$ steps and sub-dwells of $3.3$\,min for FLASH, and $K=160$ steps and sub-dwells of $3$\,min for BabyIAXO.
If each tuning move takes a time $t_m$, the segments are shortened to $K=T_{\rm sid}/(\tau+3t_m)$, and the spacing is unchanged.
}

Figure~\ref{fig:schedule} quantifies the improvement achieved with the proposed measurement schedule.
Two short sub-dwells cannot cancel both sidereal harmonics of the signal ({see Appendix~\ref{app:direction} for more details}), and a single continuous exposure reaches full coverage only after a whole sidereal day, for a generic orientation.
By contrast, three sub-dwells of equal duration spaced by $T_{\rm sid}/3$ achieve full coverage for any total exposure.

\subsection*{Orientation of FLASH}
\label{sec:strategy_flash}

For the FLASH axis orientation specified in Sec.~\ref{sec:experiments}, $n_z^2=\cos^2\alpha\cos^2\lambda\simeq0.165$, giving loss factors relative to the ideal orientation of $F_{\mathrm{excl}}=1.04$ and $F_{\mathrm{disc}}=1.18$. At the FLASH latitude, the ideal horizontal azimuth is $39.2^\circ$ from North, and the exclusion loss remains below $5\%$ for azimuths between $11^\circ$ and $59^\circ$. Since the cavity alignment is fixed by the installed solenoid, we retain this geometry and optimize the measurement schedule. Reorienting the cavity would require substantial changes to the experimental configuration and would provide only a modest additional improvement in the exclusion reach; it is therefore outside the scope of the baseline strategy considered here.

\subsection*{Pointing and modes of BabyIAXO}
\label{sec:strategy_babyiaxo}

{We emphasize here that the sensitive direction of BabyIAXO varies with the tuning of the modes, meaning that a pointing configuration optimized for a particular tuning step will not necessarily remain optimal across all scan frequencies. To identify the best overall pointing for all configurations simultaneously, we simulated $\langle c_2\rangle^\mathrm{conv}$ across all four cavities and tuning steps over the entire tuning range. A dedicated example of this analysis is shown in Fig.~\ref{fig:c2excl 4 cavities} in Appendix~\ref{app:RADES-BabyIAXO}. We first consider the case in which the plates are rotated {in the same direction.}} 
{In this scenario, when the tuning plates are kept at $0^\circ$ and the bore is horizontal,}
$\hat{\mathbf{n}}_1$ (Mode~1) is vertical, with $n_z^2=\sin^2\lambda=0.648$, and $F=1.13$ (exclusion) and $1.26$ (discovery) for any azimuth.
Raising one end of the {bore} moves $\hat{\mathbf{n}}_1$ away from the vertical: with the bore close to the North--South direction, {for an azimuth of $12^\circ$ towards East,} and its northern end raised by about $19^\circ$---within the range of the drive---$\hat{\mathbf{n}}_1$ lies at $54.7^\circ$ from the rotation axis of the Earth.
This pointing keeps Mode~1 within $0.1\%$ of the ideal orientation over the band of the first cavity. {Moreover, this alignment remains valid when the tuning plates are rotated, yielding the gain of $4.2$ in Table~\ref{tab:gain} across all tuning angles and for all four cavities simultaneously.}\\
{
For Mode~2, no single ideal configuration exists, and the sensitivity varies from one tuning step to another. The optimal pointing configuration for Mode~2 is $\alpha_b = 50^\circ$, $\beta_b = +25^\circ$, which yields mean degradation factors of $\overline{F}_\mathrm{excl} = 1.013$ and $\overline{F}_\mathrm{disc} = 1.048$ across the different tuning angles. To quantify the worst-case performance, the least favorable alignment occurs when the tuning plates are set to $0^\circ$ and $90^\circ$; in this case, $\hat{\mathbf{n}}_2$ is completely horizontal and thus independent of $\beta_b$, resulting in $F^\mathrm{worst}_\mathrm{excl} = 1.022$ and $F^\mathrm{worst}_\mathrm{disc} = 1.097$. Finally, a compromise configuration that optimizes both modes simultaneously is $\alpha_b = 46^\circ$, $\beta_b = +16^\circ$, yielding $\overline{F}_\mathrm{excl} = 1.019$ and $\overline{F}_\mathrm{disc} = 1.069$. Here too, the worst alignment occurs when Mode~2 is horizontal, giving $F^\mathrm{worst}_\mathrm{excl} = 1.032$ and $F^\mathrm{worst}_\mathrm{disc} = 1.14$. Conversely, when the plates are rotated by opposite angles, $\hat{\mathbf{n}}_1$ and $\hat{\mathbf{n}}_2$ remain fixed, and both modes stay within roughly $1\%$ of the ideal orientation when the bore end at azimuth $64.5^\circ$ from North is elevated by $25^\circ$. A summary of these results is presented in Table~\ref{tab:modes_pointing}.}
In all cases, the gain over the published projection lies between $4.0$ and $4.2$.

%%%%%%%%%%%%%%%%%%%%%%%%%%%%
%%%%%%%%%%%%%%%%%%%%%%%%%%%%

\begin{table}[t]
\caption{Optimal magnet orientation and losses $F$ for each mode choice, averaged over the tuning angles (overline) and in the worst case.\vspace{0.15cm}}
\label{tab:modes_pointing}
\begin{ruledtabular}
\begin{tabular}{lccccccc}
 Modes & tuning & $\alpha_b$ & $\beta_b$ & $\overline{F}_{\rm disc}$ & $\overline{F}_{\rm excl}$ & $F^{\rm worst}_{\rm disc}$ & $F^{\rm worst}_{\rm excl}$ \\
\hline
 1 & same  & $348^\circ$ & $+19^\circ$ & $1.000$ & $1.000$ & $1.000$ & $1.000$ \\
 2 & same  & $50^\circ$  & $+25^\circ$ & $1.048$ & $1.013$ & $1.097$ & $1.022$ \\
 1+2 & same & $46^\circ$  & $+16^\circ$ & $1.069$ & $1.019$ & $1.14$  & $1.032$ \\
 1+2 & opposite & $64.5^\circ$ & $+25^\circ$ & $1.012$&$1.003$&$1.012$&$1.003$
\end{tabular}
\end{ruledtabular}
\end{table}
%\begin{table}[t]
%\caption{Pointing of BabyIAXO for the DP run and largest loss $F$ over the band of the first cavity and over the modes read.
%The pointing is given by the azimuth from North of the raised end of the bore and by its elevation.
%Plates: tuning plates turned in the same sense (parallel) or by opposite angles~\cite{Puente:2026mog}.
%The first row neglects the rotation of $\hat{\mathbf{n}}$ by the plates, as in Ref.~\cite{Ahyoune:2023gfw}.
%\vspace{0.15cm}}
%\label{tab:babyiaxo}
%\begin{ruledtabular}
%\begin{tabular}{lllcc}
%modes & plates & pointing & $F_{\rm excl}$ & $F_{\rm disc}$ \\
%\hline
%1 & -- & horizontal & $1.13$ & $1.26$ \\
%1 & any & $\sim0^\circ$, $+19^\circ$ & $1.00$ & $1.00$ \\
%1, 2 & opposite & $64.5^\circ$, $+25^\circ$ & $1.003$ & $1.012$ \\
%1, 2 & parallel & $48^\circ$, $+15^\circ$ & $1.05$ & $1.12$ \\
%\end{tabular}
%\end{ruledtabular}
%\end{table}

%%%%%%%%%%%%%%%%%%%%%%%%%%%%
%%%%%%%%%%%%%%%%%%%%%%%%%%%%
 \begin{figure*}[t]
    \centering
    \includegraphics[width=\textwidth]{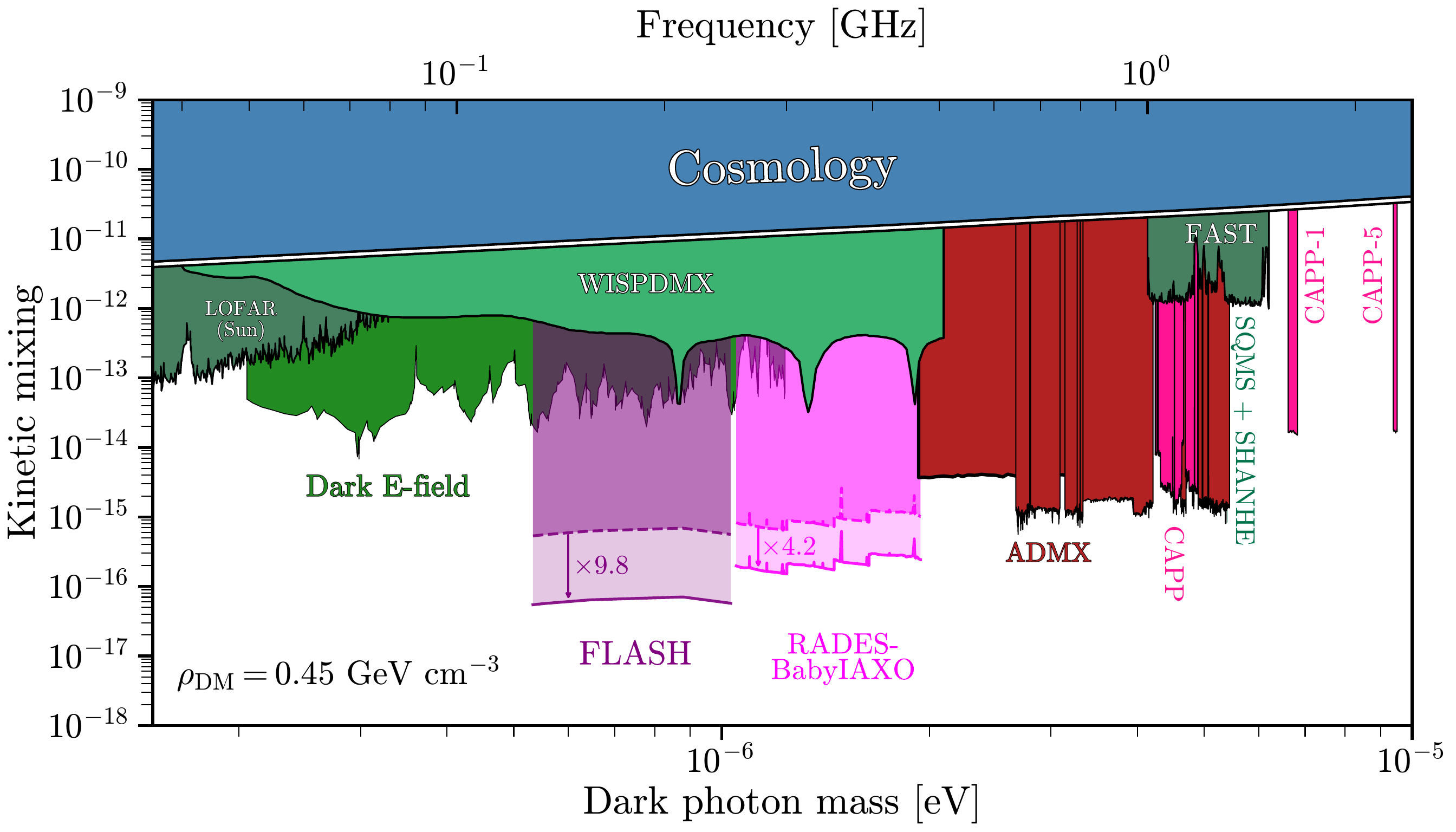}
    \caption{Sensitivity projections to polarized DP DM for FLASH and RADES-BabyIAXO (pale shade) in the $m-\chi$ plane, together with existing constraints (solid shade), assuming $\rho_\mathrm{DM} = 0.45\; \GeV/\mathrm{cm}^3$. In dashed, the published sensitivity to polarized DP DM (taking $\langle\cos^2\theta\rangle^\mathrm{excl} = 0.0025$ for FLASH and $\langle\cos^2\theta\rangle^\mathrm{excl} = 0.019$ for RADES-BabyIAXO). In solid, the sensitivity projection in the best-strategy scenario ($F_\mathrm{excl} = 1.042$ for FLASH and $F_\mathrm{excl} = 1$ for RADES-BabyIAXO's Mode 1).  Our DP limits are at 95\% C.L. The expected sensitivity to the kinetic mixing improves by about
one order of magnitude for FLASH and by a factor of four for BabyIAXO. Raw data from the repository in~\cite{AxionLimits}. 
}
    \label{fig:sensitivity_projection}
\end{figure*}
%%%%%%%%%%%%%%%%%%%%%%%%%%%%
%%%%%%%%%%%%%%%%%%%%%%%%%%%%

%%%%%%%%%%%%%%%%%%%%%%%%%%%
%%%%%%%%%%%%%%%%%%%%%%%%%%%
\section{Discussion and Conclusions}
\label{sec:conclusions}
%%%%%%%%%%%%%%%%%%%%%%%%%%%
%%%%%%%%%%%%%%%%%%%%%%%%%%%

We have established a data-taking and analysis strategy for upcoming dark-photon (DP) dark matter searches with FLASH at INFN LNF and the RADES cavities in BabyIAXO at DESY, targeting fixed, unknown polarization without {or with minimal} hardware modifications, nor disruption to axion searches~\cite{Alesini:2023qed,Ahyoune:2023gfw,Caputo:2021eaa}.
Combining noise and polarization fluctuations into a single confidence-level criterion corrects overly conservative recasts of axion projections, while splitting each frequency exposure into $N \ge 3$ equal sub-dwells spaced by $T_{\mathrm{sid}}/N$ cancels both sidereal harmonics and recovers the full sidereal-day average for any total exposure~\cite{Caputo:2021eaa}.

Together, joint statistics and a sidereal-locked scan increase the expected $95\%$~C.L.\ reach in kinetic mixing $\chi$ by a factor of $9.8$ for FLASH (with its fixed horizontal axis at azimuth $57^\circ$, $n_z^2 \simeq 0.165$, and $F_{\mathrm{excl}} = 1.042$) and by $3.7$ for BabyIAXO (Mode~1, horizontal magnet, $n_z^2 \simeq 0.648$).
During magnet-off operation~\cite{Ahyoune:2023gfw}, tilting the BabyIAXO bore to $(\alpha_b, \beta_b) \simeq (348^\circ, +19^\circ)$ places the Mode~1 sensitive direction at $\simeq 54.7^\circ$ from the Earth's rotation axis across all four scaled RADES cavities and tuning angles ($F_{\mathrm{excl}} = F_{\mathrm{disc}} = 1.000$), raising the gain over published projections to $4.2$.
Reading out the orthogonal quasi-$\mathrm{TE}_{111}$ mode (Mode~2) in parallel via a second port provides an additional DP channel and maintains a $4.0$--$4.2$ gain across opposite or parallel tuning-plate rotations with near-optimal bore pointings~\cite{Valero:2024ncz,Puente:2026mog}.

Figure~\ref{fig:sensitivity_projection} summarizes these improvements in the $(m, \chi)$ plane for $\rho_{\mathrm{DM}} = 0.45~\mathrm{GeV}/\mathrm{cm}^3$, contrasting the revised projections under our data-taking strategy (solid) against the previously published fixed-polarization projections (dashed, corresponding to $\langle\cos^2\theta\rangle^{\mathrm{excl}} = 0.0025$ for FLASH and $\langle\cos^2\theta\rangle^{\mathrm{excl}} = 0.019$ for RADES-BabyIAXO).
Implementing this strategy before data taking begins allows both FLASH and RADES-BabyIAXO to probe substantially deeper into untested dark-photon parameter space around the $\mu\mathrm{eV}$ mass scale using their baseline instrumentation.

%%%%%%%%%%%%%%%%%%%%%%%%%%%
%%%%%%%%%%%%%%%%%%%%%%%%%%%
\acknowledgments
%%%%%%%%%%%%%%%%%%%%%%%%%%%
%%%%%%%%%%%%%%%%%%%%%%%%%%%

We would like to thank Jose Reina Valero for clarifying some points of their works. The work of FM is supported
by the Italian Ministry of University and Research (MUR) via the FIS2 Consolidator Grant
project FIS-2023-02106 – QAXION (CUP: I53C25001880001). 
MG acknowledges support from project CNS2025-165965, “Señales de axiones en el rango MeV desde fuentes astrofísicas hasta detectores actuales y futuros”, funded by MICIU/AEI/10.13039/501100011033.
In addition, MG and VF acknowledge support from 
the Spanish Agencia Estatal de Investigación under grant PID2019-108122GB-C31, and from the ``European Union NextGenerationEU/PRTR'' (Planes Complementarios, Programa de Astrofísica y Física de Altas Energías). 
He also acknowledges support from grant PGC2022-126078NB-C21, ``Aún más allá de los modelos estándar".
Additionally, MG and II acknowledges funding from the European Union's Horizon 2020 research and innovation program under the European Research Council (ERC) grant agreement ERC-2017-AdG-788781 (IAXO+).
IGI and CC acknowledge funding through the European Research Council grant ERC-SYG 101118911  (DarkQuantum) and the support of Deutsche Forschungsgemeinschaft (DFG) through Grant No. 532766533 (QUANTERA) and under Germany’s Excellence Strategy – EXC 2094 – 390783311. II also acknowledges support from grant PID2022-137268NB-C51, funded by MCIN/AEI/10.13039/501100011033 and by the European Regional Development Fund (ERDF), as well as from the QuantERA project QRADES (grant PCI2024-153487), funded by MICIU/AEI/10.13039/501100011033 and co-funded by the European Union.

%%%%%%%%%%%%%%%%%%%%%%%%%%%
%%%%%%%%%%%%%%%%%%%%%%%%%%%

\appendix

\section{Time-averaged response to a fixed polarization}
\label{app:direction}

In this Appendix, we present the analytical results used in the main text in full detail. We begin by expressing all relevant quantities in the \textbf{equatorial frame}, defined by the orthonormal basis vectors $\hat{\mathbf{X}}_e$, $\hat{\mathbf{Y}}_e$, and $\hat{\mathbf{Z}}_e$. The \textbf{laboratory frame} is instead specified by the orthogonal vectors $\hat{\mathbf{N}}(t)$, $\hat{\mathbf{W}}(t)$, and $\hat{\mathbf{Z}}(t)$, pointing respectively toward the local north, west, and zenith directions. In terms of the laboratory latitude $\lambda$ and the Earth's angular velocity $\omega_\oplus=2\pi/T_{\rm sid}$, these vectors are given by
\begin{equation}
\label{eq:App-Terrestrial_coordinates}
\begin{gathered}
\hat{\mathbf{N}}(t) =
\begin{pmatrix}
-\sin \lambda \cos (\omega_{\oplus} t) \\
-\sin \lambda \sin (\omega_{\oplus} t) \\
\cos \lambda
\end{pmatrix}, \\[1em]
\hat{\mathbf{W}}(t) =
\begin{pmatrix}
\sin (\omega_{\oplus} t) \\
-\cos (\omega_{\oplus} t) \\
0
\end{pmatrix}, \qquad
\hat{\mathbf{Z}}(t) =
\begin{pmatrix}
\cos \lambda \cos (\omega_{\oplus} t) \\
\cos \lambda \sin (\omega_{\oplus} t) \\
\sin \lambda
\end{pmatrix}.
\end{gathered}
\end{equation}
We parametrize the direction $\hat{\mathbf{n}}$ in terms of a tilting angle $\beta$ with respect to the horizon and an azimuth angle $\alpha $, measured from North to West. In the lab-frame,
\begin{equation}
\hat{\mathbf{n}}(t)
= \cos \beta \left[ \cos \alpha \, \hat{\mathbf{N}}(t)
+ \sin \alpha \, \hat{\mathbf{W}}(t) \right]
+ \sin \beta \, \hat{\mathbf{Z}}(t) \,.
\end{equation}
In the notation of Ref.~\cite{Caputo:2021eaa}, $\alpha=\beta=0$ corresponds to a North-pointing detector, $\beta=90^\circ$ to a Zenith-pointing one and $(\alpha,\beta)=(90^\circ,0)$ to a West-pointing one. The equatorial components of $\hat{\mathbf n}$ are
\begin{equation}
\begin{aligned}
n_x(t)&=A\cos(\omega_\oplus t)+B\sin(\omega_\oplus t)=R\cos(\omega_\oplus t-\phi)\,,\\
n_y(t)&=A\sin(\omega_\oplus t)-B\cos(\omega_\oplus t)=R\sin(\omega_\oplus t-\phi)\,,\\
n_z&=\cos\beta\cos\alpha\cos\lambda+\sin\beta\sin\lambda\,,
\end{aligned}
\end{equation}
with
\begin{equation}
\begin{aligned}
A&=\sin\beta\cos\lambda-\cos\alpha\cos\beta\sin\lambda\,,\qquad B=\cos\beta\sin\alpha\,,\\
R&=\sqrt{A^2+B^2}=\sqrt{1-n_z^2}\,,\qquad \phi=\arctan(B/A)\,.
\end{aligned}
\end{equation}
Note that $n_z= \pm \sqrt{1-R^2}$, and the only time dependence appears in $\omega_\oplus t$. Finally, the DP polarization is expressed in terms of the angles $\theta_X,\phi_X$, 
\begin{equation}
\hat{\mathbf{X}} =
\begin{pmatrix}
\sin \theta_X \cos \phi_X \\
\sin \theta_X \sin \phi_X \\
\cos \theta_X
\end{pmatrix}.
\end{equation} 

We introduce the short notations ${s \equiv \sin \theta_X}$, ${c \equiv \cos \theta_X}$, ${\psi \equiv \phi+\phi_X}$. The projection of $\hat{\mathbf{n}}(t)$ over $\hat{\mathbf{X}}$ yields
\begin{align}
\label{eq:A-cos-theta-general-simplified}
\cos \theta(t) \equiv \hat{\mathbf{n}}\cdot\hat{\mathbf{X}}=s R \cos (\omega_\oplus t-\psi)+c n_z\,,
\end{align}  
and its square,
\begin{align}
\label{eq:App-cos2}
\cos^2\theta(t)={}&c^2n_z^2+\frac{s^2R^2}{2}
+2scRn_z\cos(\omega_\oplus t-\psi)\nonumber\\
&+\frac{s^2R^2}{2}\cos(2\omega_\oplus t-2\psi)\,
\end{align}
is a constant plus a first and a second harmonic of the sidereal frequency. The constant term is
\begin{equation}
\label{eq: A-c20}
\begin{aligned}
\langle\cos^2\theta\rangle_0
&\equiv c^2n_z^2+\frac{s^2R^2}{2}
\\
&=\frac{1-n_z^2}{2}+\frac{3n_z^2-1}{2}\,c^2\,.
\end{aligned}
\end{equation}

For a single measurement of duration $T$ starting at $t_0$, the two harmonics average to
\begin{align}
\label{eq:A-c2T-general}
&\langle\cos^2\theta\rangle_T
\equiv\frac{1}{T}\int_{t_0}^{t_0+T}\cos^2\theta(t)\,dt
=\langle\cos^2\theta\rangle_0\nonumber\\
&\quad+2scRn_z\,\operatorname{sinc}\!\left(\frac{\omega_\oplus T}{2}\right)
\cos\!\left(\omega_\oplus t_0+\frac{\omega_\oplus T}{2}-\psi\right)\nonumber\\
&\quad+\frac{s^2R^2}{2}\,\operatorname{sinc}(\omega_\oplus T)\,
\cos\left(2\omega_\oplus t_0+\omega_\oplus T-2\psi\right),
\end{align}
where $\operatorname{sinc}x=\sin x/x$. For $T$ equal to an integer number of sidereal days both sinc factors vanish and $\langle\cos^2\theta\rangle_T=\langle\cos^2\theta\rangle_0$, as found in Ref.~\cite{Caputo:2021eaa}. For $T\ll T_{\rm sid}$ both sinc factors approach unity and the measurement returns the instantaneous value $\cos^2\theta(t_0)$.

Consider now an exposure split into blocks $k$ of duration $T_k$ starting at $t_k$. For equal noise per unit time the blocks are combined with weights $w_k=T_k/\sum_jT_j$, and the average of $\cos(m\omega_\oplus t-m\psi)$ over block $k$ is ${\rm Re}\,[e^{im(\omega_\oplus t_k+\omega_\oplus T_k/2-\psi)}]\,{\rm sinc}(m\omega_\oplus T_k/2)$. Then, the combined response is
\begin{equation}
\begin{aligned}
\langle\cos^2\theta\rangle={}&\langle\cos^2\theta\rangle_0
+2scRn_z\,{\rm Re}\!\left(h_1e^{-i\psi}\right)\\
&+\frac{s^2R^2}{2}\,{\rm Re}\!\left(h_2e^{-2i\psi}\right),
\label{eq:c2_general}
\end{aligned}
\end{equation}
where the schedule enters only through the two harmonic weights
\begin{equation}
\label{eq:hm}
h_m=\sum_kw_k\,e^{im\omega_\oplus(t_k+T_k/2)}\operatorname{sinc}\!\left(\frac{m\omega_\oplus T_k}{2}\right),
\quad m=1,2,
\end{equation}
i.e.\ the averages of $e^{im\omega_\oplus t}$ over the exposure, with $|h_m|\le1$. A common shift of the start times multiplies $h_m$ by a phase, which is equivalent to a shift of $\phi_X$. For a single block, $h_m=e^{im\omega_\oplus(t_0+T/2)}\operatorname{sinc}(m\omega_\oplus T/2)$ and Eq.~\eqref{eq:c2_general} reduces to Eq.~\eqref{eq:A-c2T-general}; for $N$ blocks of equal duration and noise Eq.~\eqref{eq:c2_general} is equivalent to Eq.~\eqref{eq:avg}.

We call full sidereal coverage any schedule with $h_1=h_2=0$, for which
\begin{equation}
\label{eq:NT condition}
\langle\cos^2\theta\rangle=\langle\cos^2\theta\rangle_0
\end{equation}
for every polarization, Eq.~\eqref{eq:fullcov}. Equation~\eqref{eq:hm} shows when this is achieved.

\textit{(i) One long block.} A single block lasting an integer number $n$ of sidereal days gives $h_m\propto\operatorname{sinc}(mn\pi)=0$.

\textit{(ii) $N$ equally spaced blocks.} For $N$ blocks of equal duration $T$ and equal noise, starting at $t_k=t_0+kT_{\rm sid}/N$,
\begin{equation}
\label{eq:hm_equal}
h_m=e^{im\omega_\oplus(t_0+T/2)}\operatorname{sinc}\!\left(\frac{m\omega_\oplus T}{2}\right)
\frac{1}{N}\sum_{k=0}^{N-1}e^{2\pi imk/N}\,.
\end{equation}
The geometric sum vanishes unless $N$ divides $m$, so that $h_1=0$ for $N\ge2$ and $h_2=0$ for $N\ge3$: full coverage is reached for any $T$ if and only if $N\ge3$. The duration $T$, which only has to satisfy $T\le T_{\rm sid}/N$ for the blocks not to overlap, sets the total exposure $NT$, and hence the noise, but not $\langle\cos^2\theta\rangle$. For $N=3$, this reproduces the behavior shown in Fig.~9 of Ref.~\cite{Caputo:2021eaa}.

\textit{(iii) Two blocks.} For two blocks of duration $T$ whose start times are separated by $\Delta$,
\begin{equation}
|h_m|=\left|\operatorname{sinc}\!\left(\frac{m\omega_\oplus T}{2}\right)\cos\!\left(\frac{m\omega_\oplus\Delta}{2}\right)\right|.
\end{equation}
Canceling a harmonic requires sampling it half a period apart. The spacing $\Delta=T_{\rm sid}/2$ gives $h_1=0$ but leaves $|h_2|=|\operatorname{sinc}(\omega_\oplus T)|\simeq1$ for short blocks, while $\Delta=T_{\rm sid}/4$ gives $h_2=0$ but leaves $|h_1|\simeq1/\sqrt2$. Two blocks therefore reach full coverage only if the sinc factors help, i.e.\ if the exposure spans a whole sidereal day (e.g.\ $T=\Delta=T_{\rm sid}/2$). The exception is $n_z=0$, for which the first harmonic is absent from Eq.~\eqref{eq:c2_general} and $\Delta=T_{\rm sid}/4$ suffices, in agreement with the 6-hour spacing found in Ref.~\cite{Caputo:2021eaa} for West-pointing detectors.

\textit{(iv) Single short exposure.} A single exposure much shorter than a sidereal day has $h_m\to e^{im\omega_\oplus t_0}$, and Eq.~\eqref{eq:c2_general} returns $\langle\cos^2\theta\rangle=\cos^2\theta(t_0)$. For an isotropic prior, $\cos\theta(t_0)=\hat{\mathbf n}(t_0)\cdot\hat{\mathbf X}$ is uniformly distributed in $[-1,1]$, like $\cos\theta_X$. The response of a single short exposure, at any orientation, has therefore the same distribution as the full-coverage response with $n_z^2=1$, $\langle\cos^2\theta\rangle_0=\cos^2\theta_X$, which gives the last row of Table~\ref{tab:conv}.

With full sidereal coverage, Eq.~\eqref{eq:NT condition} holds and therefore $\langle \cos^2\theta\rangle =\langle \cos^2\theta\rangle_0 $ depends on the orientation and on the site only through $n_z^2$, and on the polarization only through $c^2$ (see Eq.~\eqref{eq: A-c20}). For $n_z^2=1/3$ one finds from Eq.~\eqref{eq: A-c20} that
\begin{equation}
    \langle\cos^2\theta\rangle_0 = \frac{1}{3}\,,
\end{equation}
independently of the orientation of $\hat{\mathbf X}$. Using Eq.~\eqref{eq:nz}, this condition directly yields the \emph{ideal-axis condition}, Eq.~\eqref{eq:ideal}, quoted in the main text.

For a general schedule, $\langle\cos^2\theta\rangle$ depends on both $c$ and $\psi$.
For an isotropic prior, $c$ is uniform in $[-1,1]$ and $\psi$ is uniform in $[0,2\pi)$ and independent of $c$.
Since Eq.~\eqref{eq:c2_general} is invariant under $(c,\psi)\to(-c,\psi+\pi)$, the condition~\eqref{eq:master} becomes
\begin{equation}
\int_0^1dc\int_0^{2\pi}\frac{d\psi}{2\pi}\,\Phi\!\left(z_{\rm obs}-\frac{P_X'}{\sigma_N}\langle\cos^2\theta\rangle\right)=1-{\rm CL}\,,
\label{eq:master_general}
\end{equation}
with $\langle\cos^2\theta\rangle$ from Eq.~\eqref{eq:c2_general}, and the conversion factor follows from Eq.~\eqref{eq:conv}.
We use Eq.~\eqref{eq:master_general} in Sec.~\ref{sec:strategy} to quantify the loss for incomplete or irregular schedules.

%%%%%%%%%%%%%%%%%%%%%%%%%%%%%%%%%%%%%%%%%%%%%%%%%%%%%%
\section{RADES-BabyIAXO: pointing optimisation for the DP search}
\label{app:RADES-BabyIAXO}
%%%%%%%%%%%%%%%%%%%%%%%%%

The RADES-BabyIAXO consists in four quasi-cylindrical  5 $\mathrm{m}$-long cavities supporting the $\mathrm{TE_{111}}$ mode. The four cavities have a fixed ratio of $b=0.9a$,
 $a$ being the width and $b$ the height. Each cavity is slightly smaller than the previous one in order to sweep different ranges of frequencies. The specific values of the four cavities are summarized in Table~\ref{tab:cav_dim}.
 \begin{table}[ht]
\small
\centering
\caption{Geometry, volume and tuning range of the four RADES-BabyIAXO cavities.  The length of all cavities is 5 m. The width of the cavity is given by $a$, and its height by $b$. The tuning plates length and thickness are 4.5~m and 10~mm, respectively.}
\vspace{2 mm}
\begin{tabular}{c c c c c c c c c c}
\hline
 Cavity & a (mm) & b (mm) &  V ($\textrm{m}^3$) & Tuning range (MHz) \\
\hline
1 & 560 & 504 &  1.195 &  252.8 - 300.3\\

2 & 492.3 & 443 & 0.876 & 300.3 - 351.6 \\

3 & 430 & 387& 0.682 & 351.2 - 395.2\\

4 & 355 & 319.5 &  0.475 & 392.2 - 469.3\\
\hline
\end{tabular}
\label{tab:cav_dim}
\end{table}
 
Individually, each cavity covers a specific frequency range. This is achieved with a mechanical tuning system based on two metallic thin rectangular plates along the longitudinal axis that turn simultaneously, as schematically shown in Fig.~\ref{fig:RADES-BabyIAXO}.
 
 The movement of the moving plates modifies the structure of the inner electromagnetic field, allowing for a frequency sweep of the mode, but also altering the direction of the sensitive directions of the modes $\mathbf{\hat{n}}$. This is shown in a 2D cut of the cavity in the xy-plane in figure \ref{fig:RADES modes}. 
 \begin{figure}[t]
     \centering
     \includegraphics[width=\linewidth]{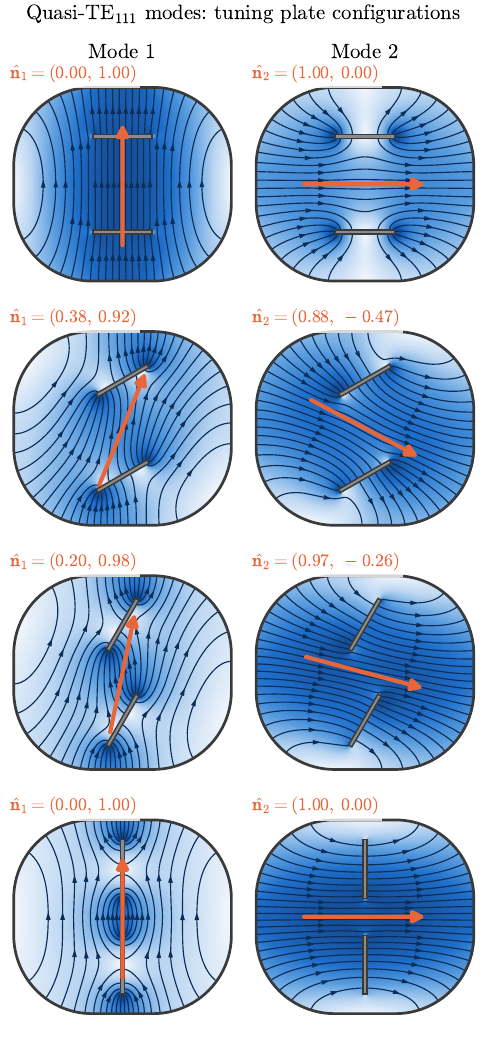}
     \caption{Electric field profiles of Mode 1 and Mode 2 in a transverse cut (xy-plane) of Cavity 1 , as well as the resulting sensitive direction $\mathbf{\hat{n}}$ for tuning plates' angles $0^\circ$, $30^\circ$, $60^\circ$ and $90^\circ$ respectively. As the plates rotate, the profile modes are modified and so do the sensitive direction of both modes.}
     \label{fig:RADES modes}
 \end{figure}
The change in $\mathbf{\hat{n}}$ implies that different frequency ranges could have different values of $\langle\cos^2\theta\rangle^\textrm{conv}$ for a given $(\alpha_b, \beta_b)$ configuration of the haloscope. In order to study the optimization of  $\langle\cos^2\theta\rangle^\textrm{conv}$ for all cavities and mode configurations, we parametrize an arbitrary vector $(n_x^c, n_y^c, n_z^c)$ in the cavity basis $\{\mathbf{\hat{x}_\mathrm{cav}}$, $\mathbf{\hat{y}_\mathrm{cav}}$, $\mathbf{\hat{z}_\mathrm{cav}}\}$, these being  the orthonormal vectors shown in fig. \ref{fig:RADES-BabyIAXO}. For a given bore orientation ($\alpha_b$, $\beta_b$), the cavity basis is expressed in the lab-frame as
\begin{equation*}
\label{eq:App-Cavity_coordinates}
\begin{aligned}
\hat{x}_{\mathrm{cav}} =
&\begin{pmatrix}
-\sin \alpha_b \\
\cos \alpha_b \\
0
\end{pmatrix},
\\[1em]
\hat{y}_{\mathrm{cav}} =
\begin{pmatrix}
-\sin \beta_b \cos \alpha_b \\
-\sin \beta_b \sin \alpha_b \\
\cos \beta_b
\end{pmatrix},&\quad 
\hat{z}_{\mathrm{cav}} =
\begin{pmatrix}
\cos \beta_b \cos \alpha_b \\
\cos \beta_b \sin \alpha_b \\
\sin \beta_b
\end{pmatrix}.
\end{aligned}
\end{equation*}
Note that for $\alpha_b = \beta_b = 0^\circ$ (horizontal magnet) the cavity basis corresponds exactly to the lab-frame. The arbitrary vector $\mathbf{\hat{n}_\mathrm{cav}}$ in the lab-frame is then
\begin{align}
n_N &= -n_x^c \sin\alpha_b - n_y^c \sin\beta_b \cos\alpha_b + n_z^c \cos\beta_b \cos\alpha_b,\nonumber \\
n_Z &= n_y^c \cos\beta_b + n_z^c \sin\beta_b\\
n_W &= \phantom{-}n_x^c \cos\alpha_b - n_y^c \sin\beta_b \sin\alpha_b + n_z^c \cos\beta_b \sin\alpha_b, \nonumber
\end{align}
 and one can obtain the ideal-axis condition for a mode with sensitive direction $\mathbf{\hat{n}_\mathrm{cav}}$ and some $(\alpha_b, \beta_b)$ using \eqref{eq:App-Terrestrial_coordinates} and requiring $n_z^2 = 1/3$ in the equatorial frame. In order to maximize $\langle \cos^2\theta\rangle^\mathrm{conv}$ for both exclusion and discovery, we have computed \eqref{eq:conv} for all four cavities and all possible configurations $(\alpha_b,\beta_b)$ for tuning plates' angles $(0^\circ,\;30^\circ,\;45^\circ,\;60^\circ,\;90^\circ)$. As an example, in Fig.~\ref{fig:c2excl 4 cavities} we show the skymaps of $\langle\cos^2\theta\rangle^\mathrm{excl}$ for Mode 1 in all four cavities for the mentioned angles. 
 \begin{figure*}[t]
    \centering
    \includegraphics[width=\textwidth]{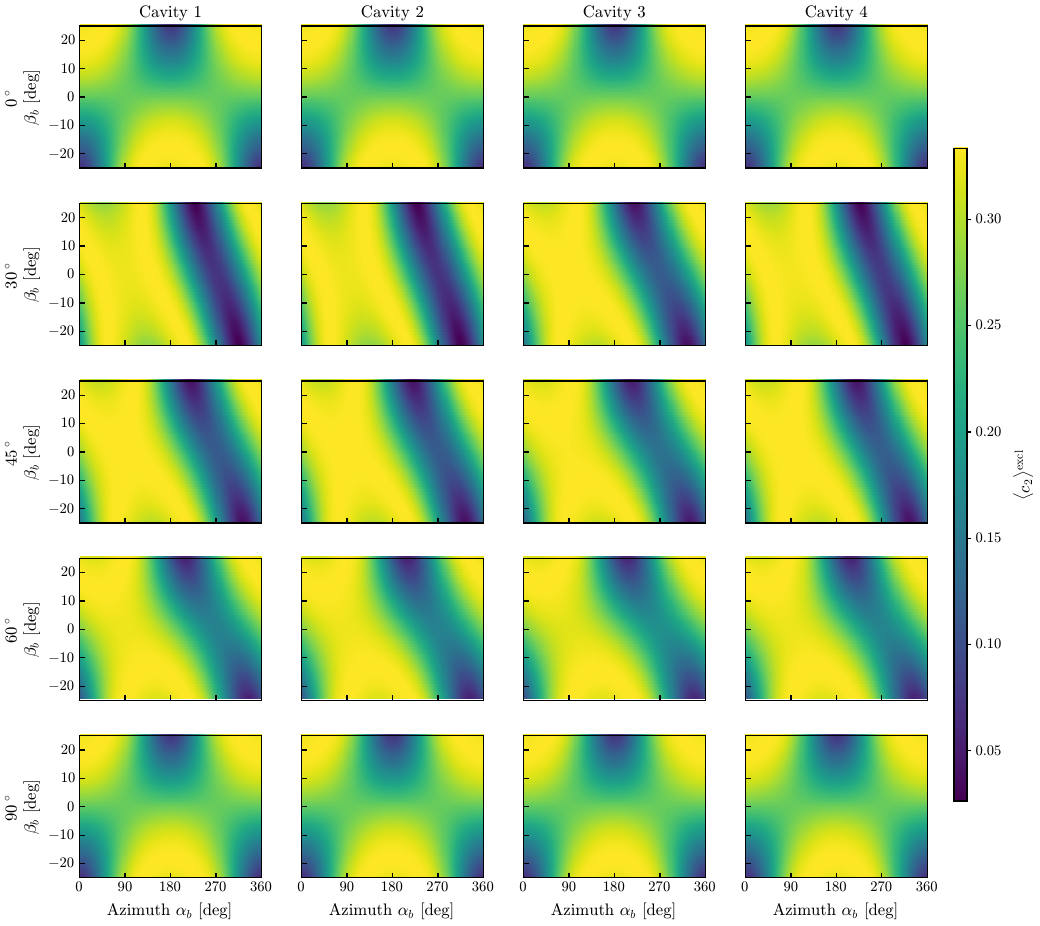}
    \caption{Skymaps of $\langle\cos^2\theta\rangle^\mathrm{excl}$ for Mode 1 in the ($\alpha_b$, $\beta_b$) parameter-space for RADES-BabyIAXO cavities 1, 2, 3, 4 (from left to right) and for tuning plates' angles $0^\circ,\;30^\circ,\;45^\circ,\;60^\circ,$ and $90^\circ$ (from top to bottom). The skymaps are almost identical from one cavity to another, allowing for a choice of ($\alpha_b$, $\beta_b$) that maximizes $\langle c_2\rangle^\mathrm{excl}$ for all cavities at the same time. }
    \label{fig:c2excl 4 cavities}
\end{figure*}
The results are almost identical from one cavity to another, which is consistent with the fact that each cavity is {similar} to one another. This simplifies the optimization, since one optimal configuration will be optimal in all cavities at once. The analysis of these simulations for both exclusion and discovery, and for both modes 1 and 2 yields the optimal configurations for $(\alpha_b,\beta_b)$ discussed in section \ref{sec:strategy_babyiaxo} for BabyIAXO and summarized in Table \ref{tab:modes_pointing}.

\bibliography{refs}
\end{document}